\documentclass[useAMS,usenatbib,a4paper]{mn2e}

\usepackage[latin1]{inputenc}

\usepackage{graphicx,epsfig}
\usepackage{util}
\usepackage{amssymb}
\usepackage[authoryear]{natbib}
\title[The fine structure of disc galaxies]{Formation of galaxies in
$\Lambda$CDM cosmologies. I. The fine structure of disc galaxies}
\author[M. Dom\'enech-Moral et al.]{M. Dom\'enech-Moral$^{1}$\thanks{E-mail:
email@address (MD); mdomenech@umh.es}, F. J. Mart\'inez-Serrano$^{1}$, R.
Dom\'inguez-Tenreiro$^{2}$ and A. Serna$^{1}$\\
$^{1}$Departamento de F\'isica y Arquitectura de Computadores, Universidad
Miguel Hern\'andez de Elche, 03202-Elche, Spain\\
$^{2}$Departamento de F\'isica Te\'orica, Universidad Aut\'onoma de Madrid,
Madrid, Spain}

\begin{document}

\date{\today}

\pagerange{\pageref{firstpage}--\pageref{lastpage}} \pubyear{2011}

\maketitle

\label{firstpage}

\begin{abstract}

In this work we present a detailed analysis of the global and fine structure of
four middle-mass disc galaxies obtained from hydrodynamic simulations in a
$\Lambda$CDM scenario. These objects have photometric disc-to-total ratios in
good agreement with those observed for late-type spirals, as well as kinematic
properties in agreement with the observational I-band Tully-Fisher relation.
We identify the different dynamical components at zero redshift on the basis of
both orbital parameters and the binding energy of stars in the galaxy. In this
way, we recognize a slowly rotating centrally concentrated spheroid, and two
disc components supported by rotation: a thin disc with stars in nearly
circular orbits, and a thick disc with orbital parameters transitional between
the thin disc and the spheroid. The spheroidal component is composed mainly by
old, metal-poor and $\alpha$-enhanced stars. The distribution of metals in this
component shows, however, a clear bimodality with a low-metallicity peak, which
could be related to a classical bulge formed from rapid collapse at early
times, and a high-metallicity peak, which could be related to a pseudo-bulge
formed from instabilities of the inner disc. The thin disc appears in our
simulations as the youngest and most metal-rich component, with median stellar
ages ranging from 3.8 to 6.7 Gyr. The radial distribution of ages and colours
in this component are U-shaped: the new stars are forming in the inner regions,
where the galaxy is bluer, and then migrate through secular processes reaching
the outer parts. Finally, we also find, in all simulated galaxies, a thick disc
containing about 16 per cent of the total stellar mass and with properties that
are intermediate between those of the thin disc and the spheroid. Its
low-metallicity stars are $\alpha$-enhanced when compared to thin disc stars of
the same metallicity. The structural parameters (e.g., the scale height) of the
simulated thick discs suggest that such a component could result from the
combination of different thickening mechanisms that include merger-driven
processes, but also long-lived internal perturbations of the thin disc.

\end{abstract}


\section{Introduction}\label{sec:Intro}

Spiral galaxies have long been recognized to contain two main stellar
components: a dynamically hot spheroidal component composed by old and
metal-poor stars, and a cold disc of young and more metal-rich stars.

In the past three decades, a wide range of observations has revealed the
existence in almost all resolved spiral galaxies of yet another component, a
thick stellar disc. This component was originally detected in the Milky Way
using star counts \citep{1983MNRAS.202.1025G} and as an excess red flux at
large galactic latitudes in external galaxies \citep{2002AJ....124.1328D,
2006AJ....131..226Y}. It is now known to be an ubiquitous component of spiral
galaxies \citep{2002AJ....124.1328D, 2005AJ....130.1574S}, but for many years
it remained unclear whether it was a truly distinct component of the Milky Way
thin disc. Evidences from their distinct nature came from detailed kinematic
and chemical abundance studies \citep[e.g.][]{1998A&A...338..161F,
2003A&A...398..141S, 2003A&A...410..527B, 2005A&A...433..185B,
2006MNRAS.367.1329R}.

In general thick discs are characterized by an exponential-fit scale height,
$z_h$, larger than the thin disc by a factor of $\sim$2
\citep[e.g.][]{2002AJ....124.1328D, 2006AJ....131..226Y}. In our Galaxy and in
NGC 891, a Milky Way analogue, the $z_h$ value seems to be roughly constant
with radius \citep{2003A&A...409..523R, 2009MNRAS.395..126I}. Relative to the
galactic thin disc, thick disc stars have both larger velocity dispersions and
slower net rotation \citep{2003A&A...398..141S}. It contains 10 to 35 per cent
of the total baryonic mass of a galaxy \citep{2006AJ....131..226Y,
2010arXiv1010.5276C}, and appears to be anti correlated with the bulge mass:
galaxies with small bulges have larger thick discs. Concerning their stellar
population, thick disc stars are older and more metal-poor than stars in the
thin disc \citep[e.g.][]{1995AJ....109.1095G, 1998A&A...338..161F}. They are
also significantly enhanced in $\alpha$-elements compared to thin disc stars of
comparable iron abundance \citep[e.g.][]{1998A&A...338..161F,
2003A&A...410..527B, 2005A&A...433..185B, 2006MNRAS.367.1329R}, suggesting a
star formation time-scale shorter than 1 Gyr \citep{1998A&A...338..161F,
2010ApJ...721L..92R}.

Several formation mechanisms have been proposed in the literature to explain
the origin of the different galactic components. In the prevailing picture of
galaxy formation, haloes of dark matter acquire angular momentum via tidal
torques from interacting structures. As gas cools and condenses into their
central parts, star-forming galaxies appear with a thin disc structure
\citep{1980MNRAS.193..189F, 1984ApJ...286...38W}. Hydrodynamical simulations,
however, have shown that cold flows play a fundamental role in the building up
of discs in low-mass haloes \citep[e.g.][]{2005MNRAS.363....2K,
2006MNRAS.368....2D}. The cold gas flows rapidly to the centre of galaxies from
filamentary structures around haloes, where it may be swiftly accreted into the
central galaxy. The lower initial gas temperature leads to shorter cooling
times and to the growth of stellar discs at higher redshifts than predicted by
the standard model \citep{2009ApJ...694..396B}.

Within these pictures, two main formation scenarios have been proposed for the
bulges of spiral galaxies \citep[see][for a review]{2004ARA&A..42..603K}. In
classical bulges, most stars are formed during the early phase of intensive
star formation following collapse of the proto-galaxy and subsequent mergers.
In another scenario, boxy or pseudo-bulges can also be developed through
dynamical instability of the inner disc.

Thin disc instabilities are also considered as the main responsible for the
thick disc formation. In most of the proposed scenarios the thick disc stars
are formed in the thin disc and are dynamically heated through either the
accretion of galaxy satellites \citep{1993ApJ...403...74Q, 1996ApJ...460..121W,
2008MNRAS.391.1806V} or internal perturbations of the thin disc from long-lived
mechanisms \citep[e.g., the stellar scattering via resonant interactions with
spiral arms][]{2008ApJ...684L..79R,2009MNRAS.396..203S,2009MNRAS.399.1145S} or
short-lived events \citep[e.g., perturbations due to the thin disc clumpy
stellar structure at high redshifts][]{2009ApJ...707L...1B}. In other category
of models, the thick disc stars are assumed to be formed either {\it in situ}
during/after gas-rich mergers \citep{2004ApJ...612..894B,2005ApJ...630..298B},
or in external satellites that are disrupted and incorporated into the thick
disc \citep{2003ApJ...597...21A}.

Numerical hydrodynamic simulations have been recognized as a powerful tool to
study galaxy formation and evolution in its full cosmological context. The main
difficulty in reproducing realistic disc-dominated galaxies is the transfer of
gas angular momentum to the dark matter \citep{1994MNRAS.267..401N}, which
results in small simulated discs with too high bulge-to-disc ratios. This
catastrophic loss of angular momentum can be prevented by modelling processes
(e.g., the feedback from supernova explosions) that regulate star formation and
the rapid consumption of gas at early times. Although the observed large
fraction of late type galaxies is still far from being reproduced in
simulations \citep{2009ApJ...696..411W}, in the past decade some simulations
have been able to reproduce ensemble properties of disc galaxies with gaseous
discs at lower resolutions \citep{1998ApJ...508L.123D, 2001MNRAS.325..119S}, as
well as to produce individual examples of realistic disc galaxies in
$\Lambda$CDM universes \citep[e.g.][]{2003ApJ...596...47S, 2003ApJ...597...21A,
2004ApJ...612..894B, 2007MNRAS.374.1479G, 2009MNRAS.398..312G,
2008MNRAS.389.1137S}. These simulations have generated extended stellar disc
components with high specific angular momentum and with a significant fraction
of young stars \citep{2008MNRAS.389.1137S}. Although an old and significant
spheroid is almost always present \citep{2003ApJ...596...47S,
2003ApJ...591..499A, 2007MNRAS.374.1479G, 2008MNRAS.389.1137S}, these discs
show realistic exponential scale lengths and integrated colours. Moreover,
they fall on the region defined by the observational I-band Tully-Fisher
relation and have bulge-to-disc ratios or stellar mass-to-light ratios similar
to observational counterparts. The presence of a thick disc component is
evidenced in some simulated discs by both a double exponential vertical
structure, and a velocity dispersion versus age relation that resemble those
observed in the Milky Way \citep{2003ApJ...597...21A, 2004ApJ...612..894B}.

The aim of this and forthcoming papers is to study the formation mechanisms of
the different galactic components (bulge, halo, thin disc and thick disc) by
means of cosmo-hydrodynamic simulations. For this purpose we follow the
evolution of four different disc galaxies in a $\Lambda$CDM cosmology, using an
OpenMP parallel version of the DEVA code \citep{2003ApJ...597..878S} which
includes the chemical feedback and cooling methods described in
\citet{2008MNRAS.388...39M}. The present work is focussed on their global and
fine structure at $z=0$ and their comparison with observations. We will show in
this paper that the dynamical and chemical properties of these simulated
galaxies have values consistent with observations and contain valuable
information about their formation processes.

This paper is organized as follows. In Section 2 we briefly describe the
simulation code and the main characteristics of our simulations. In Section 3
we present the general properties of our simulated galaxies at $z=0$ and
compare with observational disc properties. In Section 4 we identify the
dynamical components in the simulated galaxies (subsection 4.1), to perform a
detailed description and comparison of these components with observational
references. We analyse the vertical density profiles (Section 5), stellar ages
(subsection 6.1), age and colour gradients (subsection 6.2) and chemical
properties (Section 7) of such components. Finally, in Section 8 we summarise
and discuss our main results.

\section{The simulations}\label{sec:DEVA}

All simulations were performed with an OpenMP parallel version of DEVA, an
AP3M+SPH code specially designed so that conservation laws (e.g. momentum,
energy, angular momentum and entropy) hold as accurately as possible
\citep[see][for details]{2003ApJ...597..878S}. The code also includes methods
to account for subresolution physical processes such as star formation, gas
restitution from stars, chemical feedback and cooling. These methods are
extensively described in \citet{2008MNRAS.388...39M}. We will summarize them
below.

\subsection{Star formation}

Star formation (SF) is implemented through a Kennicutt-Schmidt-like law:
\begin{equation}
\frac{d\rho}{dt}=-\frac{c_* \rho}{t_g}
\end{equation}
where $\rho$ is the gas density, $c_*$ is a dimensionless SF efficiency
parameter, and $t_g$ is a characteristic time-scale chosen to be equal to the
maximum of the local dynamical time and the local cooling time. We consider
that gas particles are eligible to form stars if they are located in a region
with a convergent flow and a gas density higher than a density threshold
$\rho_*$. At each time-step we compute the probability that those eligible gas
particles form stars in the considered time interval, and draw random numbers
to decide which of them will form stars. Then, each of these randomly selected
gas particles is transformed into one stellar particle, which is assumed to be
a single stellar population (SSP) of a given age and mass (see Table
\ref{table1}). In the simulations presented in this paper, stellar masses
within each SSP were assumed to be distributed according to the Salpeter
initial mass function (IMF) \citep{1955ApJ...121..161S}, with a mass range of
[M$_l$,M$_u$]=[0.1,100]M$_\odot$.

Since our cosmological simulations do not resolve star-forming molecular
clouds, in the above scheme the choice of the $\rho_*$ value is based on the
fact that the ISM can be represented by a lognormal density probability
distribution function (PDF) \citep[see, e.g.,
][]{2003ApJ...590L...1K,2007ApJ...660..276W}. The amount of gas eligible for
star formation is represented by the high density part of the PDF, which in
turn is a function of the total galaxy gas mass and turbulence. The choice of
the $\rho_*$ value must allow for the high-density star-forming part of the PDF
to be well resolved or, at least, to contain a converged amount of star-forming
mass for the adopted numerical resolution \citep{2011MNRAS.410.1391A}. Despite
this theoretical dependence of the SF parameters on the local properties, and
probably also on redshift, for a given simulation such parameters are usually
taken as constants.

Discrete energy injection processes have been found to play an important role
in the self-regulation of the SF processes in galaxies. Positive feedback
effects can help to halt the overproduction of stellar mass and, when
implemented in simulations, help to properly reproduce some properties of disc
galaxies. The most widely employed model in cosmological hydrodynamical
simulations assumes a strong explicit feedback combined with a high SF
efficiency (\citealp[e.g.][]{2008MNRAS.389.1137S,2011arXiv1112.0315S}; but see
also the discussion in \citealp{2009MNRAS.396.2332K}). However, a number of
authors \citep[see][for a recent work]{2011MNRAS.410.1391A} have shown that the
sub-parsec processes of SF self-regulation in simulated disc galaxies can also
be implicitly modelled by means of a low SF efficiency. Since our aim in this
paper is to test the minimal conditions for the formation of realistic discs,
we have adopted the latter model for subresolution physics. According with this
approach, we have chosen inefficient SF parameters (see Table \ref{table1}),
which implicitly account for the regulation of star formation by discrete
energy injection processes.

\subsection{Chemical feedback and Cooling}

Methods for chemical feedback and cooling are also those described in
\citet{2008MNRAS.388...39M}. These methods account for the full dependence of
radiative cooling on the detailed composition of the gas, through a fast
algorithm based on a metallicity parameter, $\zeta(T)$, which takes into
account the weight of the different elements on the total cooling function.
Such a method for cooling is important to prevent an overestimation of the
metallicity-dependent cooling rate.

The release to the ISM of newly produced elements, as by-products of stellar
evolution and death, is included in our code through a stochastic algorithm
that completely removes the assumption of instantaneous recycling of stellar
ejections. In this algorithm, the metal ejection from stars is progressively
released to the ISM as increments in the metal content of nearby gas particles.
The delayed gas restitution from stars is also taken into account by means of a
probabilistic approach that reduces the statistical noise and, hence, that
allows for the study of the inner chemical structure of objects with moderately
high numbers of particles. Furthermore, once the stellar production of metals
has been released to the ISM, the chemical elements are distributed and mixed
through the gas component by means of a diffusion law that models, without the
explicit inclusion of stellar winds, the subresolution turbulent mixing of
metals.

In the computation of the metal production, our code accounts for the full
dependence on the detailed chemical composition of stellar particles by means
of the Q$_{ij}$ formalism \citep{1973ApJ...186...51T}, which relates each
nucleosynthetic product to its sources. Such a Q$_{ij}$ formalism is important
to prevent a significant underestimation of the [$\alpha$/Fe] ratio in
simulated galaxies. In this paper, we have used the stellar evolution data
described in \citet{2005A&A...432..861G} for low and intermediate mass stars,
and \citet{1995ApJS..101..181W} for high-mass stars. SNIa rates were computed
according to \citet{2000MmSAI..71..435R} and their element production according
to \citet{1999ApJS..125..439I}. We have considered the evolution of the
following elements: H, $^4$He, $^{12}$C, $^{13}$C, $^{14}$N, $^{16}$O,
$^{20}$Ne, $^{24}$Mg, $^{28}$Si, $^{32}$S, $^{40}$Ca and $^{56}$Fe.

\begin{figure*}
\begin{center}
\includegraphics[width=1.0\textwidth]{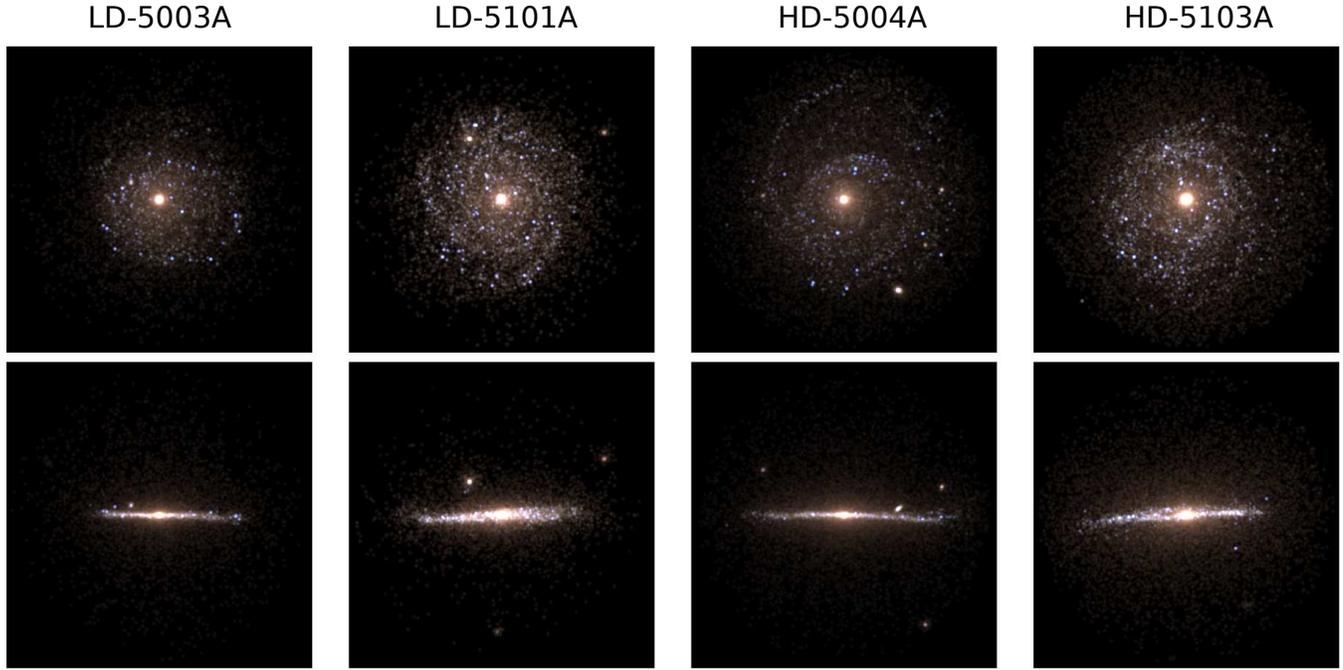}
\end{center}
\caption{Face-on and edge-on synthetic images obtained using
\citet{2003MNRAS.344.1000B} models at $z=0$. All images are 40 kpc side. }
\label{objects}
\end{figure*}

\begin{table*}
\centering
\caption{Principal characteristics of simulations: mass of a single baryonic
particle (m$_{bar}$), star-formation parameters, density threshold ($\rho_*$)
and efficiency (c$_*$), virial radius (r$_\mathrm{vir}$), spin parameter of the
dark matter halo ($\lambda_\mathrm{DM}$), the total (M$_\mathrm{vir}$), stellar
(M$_\mathrm{star}^\mathrm{vir}$) and gas (M$_\mathrm{gas}^\mathrm{vir}$) masses
inside the virial radius. \label{table1}}

\begin{tabular}{l ccccccccc}
\hline
Galaxy & m$_\mathrm{bar}$ & $\rho_*$ & c$_*$ & r$_\mathrm{vir}$ &
$\lambda_\mathrm{DM}$ & M$_\mathrm{vir}$ & M$_\mathrm{star}^\mathrm{vir}$ &
M$_\mathrm{gas}^\mathrm{vir}$ \\
& [M$_{\odot}$] & [g cm$^{-3}$] & & [$h^{-1}$ kpc] && [10$^{11}$ M$_{\odot}$] &
[10$^{10}$ M$_{\odot}$] & [10$^{10}$ M$_{\odot}$]\\
\hline
LD-5003A & 3.82$\times10^5$ & 6$\times10^{-26}$ & 0.010 & 106.12 & 0.019720 & 1.95 & 1.79 & 0.63\\
LD-5101A & 3.79$\times10^5$ & 6$\times10^{-26}$ & 0.010 &  90.55 & 0.034233 & 1.21 & 1.49 & 0.48\\
HD-5004A & 3.94$\times10^5$ & 6$\times10^{-26}$ & 0.010 & 127.00 & 0.023273 & 3.35 & 3.74 & 1.02\\
HD-5103A & 3.78$\times10^5$ & 6$\times10^{-26}$ & 0.010 & 141.62 & 0.098032 & 4.66 & 3.42 & 0.88\\
HD-5004B & 3.94$\times10^5$ & 1$\times10^{-25}$ & 0.008 & 128.10 & 0.023387 & 3.44 & 3.57 & 1.31\\
HD-5103B & 3.78$\times10^5$ & 1$\times10^{-25}$ & 0.008 & 140.53 & 0.096513 & 4.55 & 3.36 & 0.89\\
HD-5004L & 3.12$\times10^6$ & 3$\times10^{-26}$ & 0.010 & 128.47 & 0.027249 & 3.48 & 3.87 & 0.96\\
\hline
\end{tabular}
\end{table*}

\subsection{Simulated disc galaxies}

The simulation of each disc galaxy consists of a cosmological zoom-in that
includes high-resolution gas and dark matter for the flow converging region
that generates the main object. The rest of the simulation box is sampled by
low-resolution dark matter particles that account for tidal forces over the
flow converging region.

As a first step, we performed two low-resolution simulations that consisted of
two different Montecarlo realisations of the same cosmological model: a flat
\textit{$\Lambda$CDM} universe (with $\Omega_{\Lambda}=0.723$,
$\Omega_m=0.277$, $\Omega_b=0.04$, $\sigma_8 = 0.811$ and $h=0.7$) within a
periodic box of 10 Mpc per side\footnote{Although this box side implies a lack
of very massive objects and galaxy groups, it has been proved that it has
little effect on the internal properties of the haloes
\citep{2006MNRAS.370..691P} and on the formation of individual galaxies.}.

From each of these low-resolution simulations, we selected two gas-rich objects
with a prominent gas disc at $z=0$ not apparently perturbed from recent major
mergers: the most massive one from a sparsely populated region and the most
massive one from a crowded region. These selection criteria then focus on
massive regular spiral galaxies extracted from both low and high density
regions and, hence, with different histories of mergers and accretion events.
For each selected object, we traced back the particles inside its virial radius
until the initial redshift $z_\mathrm{init} = 31$. We then computed the convex
hull \citep{barber96quickhull} enclosing these particles at $z_\mathrm{init}$
and substituted all the particles inside the convex hull by their high
resolution counterparts. Gas particles outside the hull were eliminated and
their masses added to the low-resolution dark matter component, thus obtaining
the initial conditions for each high-resolution simulation. The galaxies
obtained in this way have a full history of mergers and accretion events in a
cosmological context, without any assumptions made for their initial conditions
beyond the cosmology and the initial conditions generator used
\citep{2008ApJS..178..179P}. Table \ref{table1} summarises the main
characteristics of each high-resolution simulation, with a gravitational
softening of $\epsilon_g = 400$ $h^{-1}$ pc and a minimum hydrodynamical
smoothing length half this value. In this table, LD-5003A and LD-5101A
correspond to objects selected in low density (LD) regions, while HD-5004A and
HD-5103A correspond to objects from high density (HD) regions. All these
objects (labelled as A) were simulated with identical SF parameters. The two
objects selected from crowded regions experienced major mergers at moderately
low redshifts. In principle, the ability of their discs to regenerate after
their last major merger, as well as the resulting properties of their fine
structure, could be sensitive to the choice for SF efficiency. In order to test
the possible effects of SF efficiency, these two objects were re-simulated with
less efficient SF parameters (simulations labelled as B: HD-5004B and
HD-5103B).

Table \ref{table1} also shows the total, stellar and gas masses enclosed by the
virial radius of each object. We can see from this table that our simulated
galaxies have virial masses in the range 1.2 to 4.7$\times 10^{11}$ M$_{\odot}$
and stellar masses of $\sim 10^{10.5}$, placing them among middle-mass spiral
galaxies. Therefore, they are less massive than the Milky Way (with virial mass
of $\sim$1.2$\times 10^{12}$ M$_{\odot}$) and their masses are comparable to
the least massive simulated galaxies presented in recent works
\citep[e.g.][]{2007MNRAS.374.1479G,2009MNRAS.396..696S,2011arXiv1103.6030G}.

\begin{table*}
\centering

\caption{Face-on photometric decomposition of simulated galaxies in different
bands. r$_e$ stands for the effective radius of the bulge, r$_s$ for the scale
length of the disc. L$_T$ and L$_D$ represent the luminosity in the central
region and the luminosity of the disc. We also show the $\chi^2$ value of the
fit. D/T$^P$ is defined as the disc-to-total ratio obtained using the
photometric decomposition, and M as the absolute magnitude in each band.}
\begin{tabular}{l ccccccccccccc}
\hline \hline
Galaxy & Band & n & r$_e$ & r$_s$ & L$_T$ & L$_D$ & $\chi^2$ & D/T$^{P}$ & M \\
&&& [kpc] & [kpc] & mag $arcsec^{-2}$ & mag $arcsec^{-2}$ & $\times10^{-5}$ & &
mag \\
\hline \hline
       & r & 0.78 & 0.28 & 2.72 & 18.0 & 17.0 & 6.0 & 0.72 & -20.20 \\
LD-5003A & g & 0.63 & 0.27 & 2.86 & 18.7 & 17.5 & 9.7 & 0.75 & -19.69 \\
       & i & 0.83 & 0.29 & 2.67 & 17.7 & 16.8 & 5.3 & 0.70 & -20.48 \\
\hline
       & r & 0.90 & 0.32 & 3.83 & 18.6 & 16.8 & 14.3 & 0.84 & -20.03 \\
LD-5101A & g & 0.73 & 0.31 & 4.23 & 19.2 & 17.1 & 21.3 & 0.87 & -19.58 \\
       & i & 0.96 & 0.32 & 3.76 & 18.3 & 16.6 & 12.0 & 0.83 & -20.28 \\
\hline
       & r & 2.21 & 0.37 & 4.00 & 17.3 & 16.4 & 7.7 & 0.70 & -20.83 \\
HD-5004A & g & 1.70 & 0.33 & 4.05 & 18.0 & 16.8 & 7.3 & 0.76 & -20.29 \\
       & i & 2.38 & 0.39 & 3.98 & 16.9 & 16.1 & 7.7 & 0.68 & -21.12 \\
\hline
       & r & 0.73 & 0.39 & 3.85 & 17.3 & 16.4 & 5.1 & 0.69 & -20.77 \\
HD-5103A & g & 0.58 & 0.37 & 3.83 & 18.0 & 16.8 & 9.1 & 0.75 & -20.26 \\
       & i & 0.78 & 0.40 & 3.85 & 16.9 & 16.1 & 5.2 & 0.68 & -21.05 \\
\hline
       & r & 0.80 & 0.26 & 3.29 & 17.6 & 16.2 & 9.0 & 0.77 & -20.88 \\
HD-5004B & g & 0.58 & 0.25 & 3.47 & 18.3 & 16.7 & 17.0 & 0.81 & -20.37 \\
       & i & 0.88 & 0.26 & 3.21 & 17.3 & 16.0 & 7.3 & 0.76 & -21.16 \\
\hline
       & r & 1.75 & 0.55 & 3.90 & 17.0 & 16.6 & 6.9 & 0.58 & -20.77 \\
HD-5103B & g & 1.63 & 0.48 & 3.65 & 17.6 & 17.0 & 6.4 & 0.64 & -20.24 \\
       & i & 1.76 & 0.57 & 3.99 & 16.6 & 16.3 & 7.3 & 0.56 & -21.05 \\
\hline
       & r & 1.01 & 0.31 & 3.76 & 17.3 & 16.2 & 5.0 & 0.72 & -20.87 \\
HD-5004L & g & 1.07 & 0.30 & 4.02 & 17.9 & 16.8 & 6.0 & 0.73 & -20.31 \\
       & i & 0.98 & 0.31 & 3.67 & 17.0 & 16.0 & 5.0 & 0.72 & -21.17 \\
\hline
\end{tabular}
\label{table2}
\end{table*}

\section{General properties of discs}

\subsection{Bulge-disc decomposition: D/T ratios}

Fig. \ref{objects} shows the face-on and edge-on synthetic images of the four
main simulated objects at $z=0$, where SSP ages, masses and metallicities have
been converted into luminosities using the \citet{2003MNRAS.344.1000B}
models\footnote{Dust effects have not been considered in this work, so, in
order to minimize these effects, photometric studies have been done with
galaxies seen face-on.}. In all these images, a conspicuous disc component can
be appreciated even for the object HD-5103A, where the disc structure has
survived a recent major merger at $z\sim 0.3$. Although the synthetic image of
HD-5103A does not show a clear evidence of such a merger, we will see in
Section \ref{sec:fineST} that this event has still a strong imprint in several
physical properties.

\begin{figure}
\begin{center}
\includegraphics[width=1.0\columnwidth]{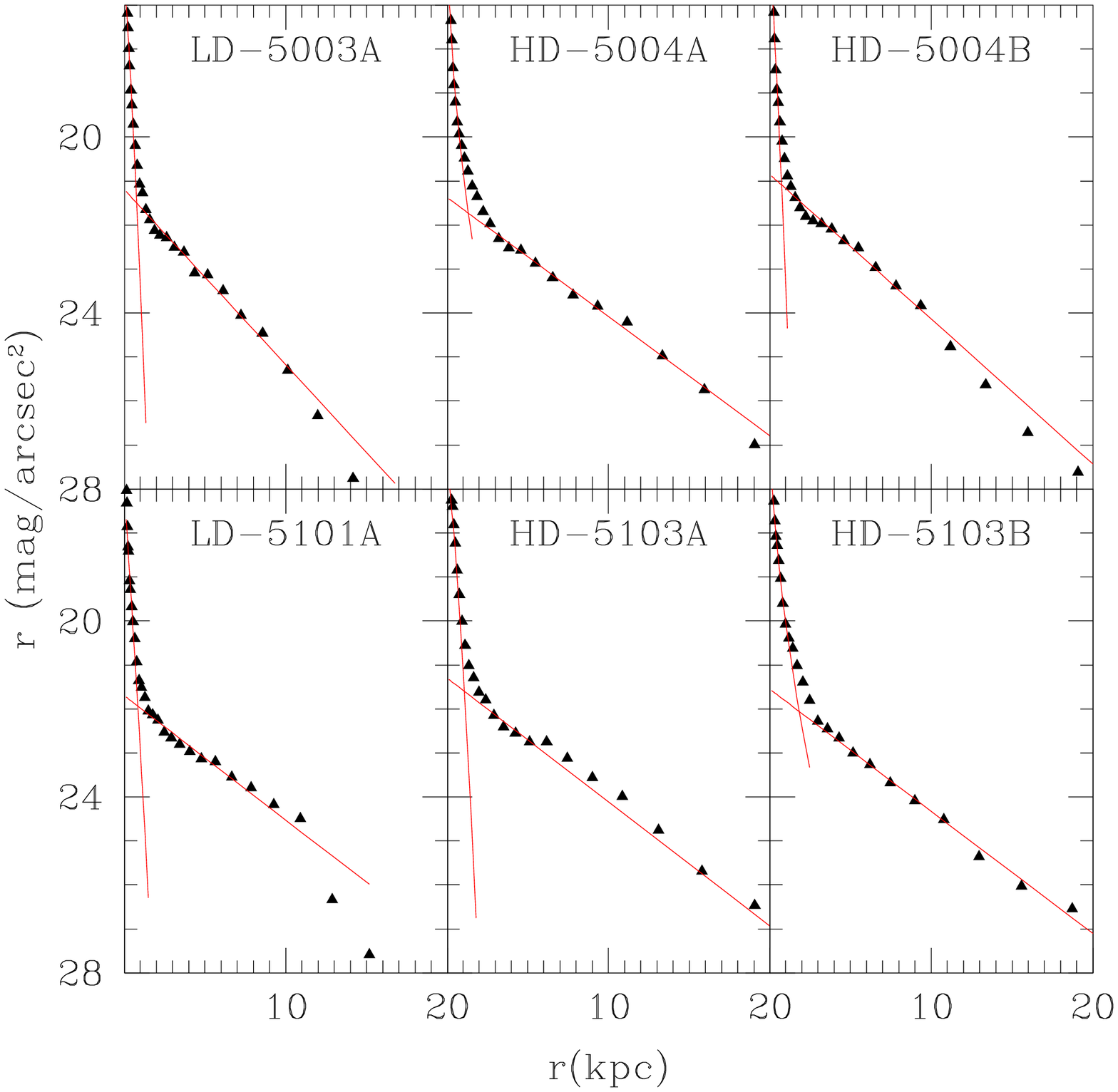}
\end{center}
\caption{Face-on luminosity profiles in the $r$ band (filled triangles) and
their fits obtained with the bulge-disc decomposition (red solid lines).}
\label{bulge-disk}
\end{figure}

In order to quantify the importance of each component, Fig. \ref{bulge-disk},
shows the face-on $r$-band luminosity profiles as well as the fits obtained
from their bulge-disc decomposition (see Table \ref{table2} for the fitting
parameters in different bands). In each band we have performed a one
dimensional fit which combines a S\'ersic law for the inner region and a simple
exponential profile in the outer parts. This fit provides a very good
description of the stellar luminosity profile up to a limit of $\sim$24 mag
$\mathrm{arcsec}^{-2}$. In the outer regions, where a second exponential should
be employed \citep[see][]{2009ApJ...705L.133M}, some deviation is instead
observed. Such a deviation has however little effect on the global fit quality
measured with the mean square deviation, $\chi^2$.

Recent works have used 2D fitting packages developed for use in observational
studies \citep[e.g.][]{2008MNRAS.384..420G} to perform the bulge-disc
decomposition. They have shown that different results can be obtained when
different fitting techniques are employed, and also that the effects of
neglecting bars when they are present can alter the resulting bulge-to-total
and disc-to-total ratios. In order to test the effects of such different
techniques, we have also performed fits to face-on synthetic images using
GALFIT \citep{2010AJ....139.2097P}. The resulting fits show similar values for
the scale length and effective radius, somewhat higher values for the S\'ersic
shape parameter $n$, and minor-to-major ratios very close to one. The resulting
D/T ratios were roughly the same as those obtained from 1D fits, but with a
higher uncertainty due to the additional parameters (point-spread function,
dust processing,...) required to generate the synthetic images. Due to the
similitude of results and the fact that our objects have a high azimuthal
symmetry, we have chosen to show here only the one-dimensional bulge-disc
decompositions.

The D/T ratios, measured in our objects from their 1D fits in different
photometric bands, are shown in Table \ref{table2}. In general, they are in
good agreement with those expected for spiral galaxies
\citep[e.g.][]{2007MNRAS.381..401L, 2009ApJ...705..245O} and are comparable
with those recently obtained in simulations
\citep[e.g.][]{2009MNRAS.398..312G,2011arXiv1103.6030G}. Further, the values
obtained from S\'ersic fits to the total light distribution, are also in good
correspondence with both the luminosity-size and stellar mass-size relations
for disc galaxies given by \citet{2003MNRAS.343..978S}. In order to compare our
results with the compilation reported by \citet{2008MNRAS.388.1708G}, we have
also computed the bulge-disc decomposition in the K-band. The median values
obtained for the shape parameter (n=1.46), the effective radius (r$_e=0.37$
kpc) and the scale length (r$_s=3.71$ kpc) of our discs are in good agreement
with those reported for spiral galaxies.

Although these results must be taken with caution, due to uncertainties
introduced by different fitting \citep{2008MNRAS.384..420G} and decomposition
\citep[e.g.][see also the Section \ref{sec:kk-means}
below]{2003ApJ...597...21A,2010MNRAS.407L..41S} methods, the relatively small
size of bulges obtained in these simulations is an important result by itself,
given the usual problems in reproducing realistic disc galaxies with
cosmological simulations \citep{2003ApJ...591..499A,2009arXiv0908.1409B}. As
already quoted in Section \ref{sec:Intro}, the main problem hindering the
formation of large galactic discs is the loss of gas angular momentum at early
stages of galaxy formation, which results in objects with too high
bulge-to-disc ratios. This catastrophic loss of gas angular momentum can be
prevented by including feedback processes avoiding the early in-fall of gas and
its rapid consumption \citep[e.g.][]{2003ApJ...596...47S, 2007MNRAS.374.1479G,
2008MNRAS.389.1137S, 2011MNRAS.410.1391A}. The cosmological simulations
reported in this work, where energy feedback is implicitly included in the star
formation parameters, lead to disc galaxies with a bulge component comparable
to those observed in Sab-Sb galaxies \citep{2009ApJ...705..245O}.

\subsection{Rotation Curves and the Tully-Fisher relation}

\begin{figure}
\includegraphics[width=1.00\columnwidth]{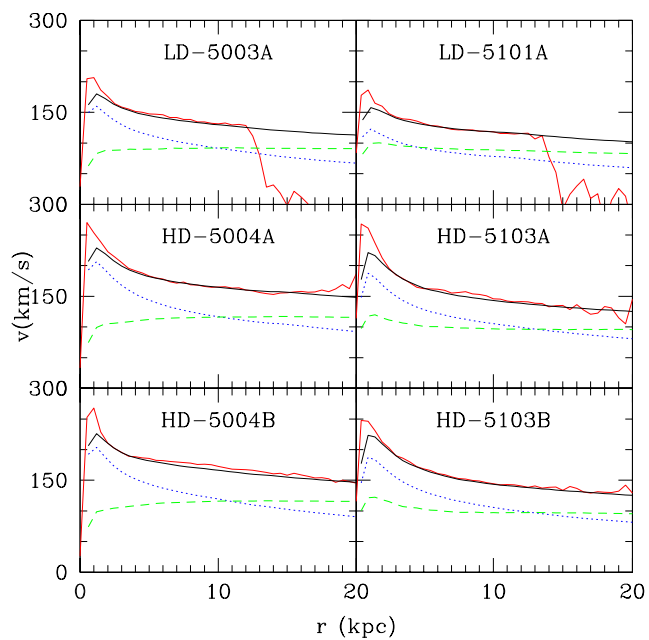}
\caption{Circular velocity profiles of the simulated galaxies. Black solid
line stands for the total mass distribution ($V_\mathrm{circ}$), as well as
green dashed and blue dotted lines for the dark matter ($V_\mathrm{dm}$) and
baryonic ($V_\mathrm{bar}$) contributions, respectively. The rotational mean
velocities for the gas particles are displayed with a red solid line.}
\label{crot}
\end{figure}

Fig. \ref{crot} shows the circular velocity curves for the dark matter and the
baryonic components (dashed and dotted lines, respectively) of the simulated
galaxies, computed as
\begin{equation}
V_i^2(r)=\frac{GM_i(<r)r^2}{(r^2+\epsilon_g^2)^{3/2}}
\end{equation}
with $i$=baryons (bar) and dark matter (dm). Solid lines are the circular
velocities, $V^2_\mathrm{circ}(r)=V^2_\mathrm{dm}(r)+V^2_\mathrm{bar}(r)$,
where r stands for the radial distance to the galactic centre. As it can be
clearly seen from this figure, the baryonic component of the galaxy dominates
the dynamical properties in the innermost region, while the contribution of
dark matter is approximately constant across the galaxy, dominating the curve
in the external region. The red solid line stands for the rotational velocity
measured from the gas particle movement. We can see that gas particles closely
follow the circular velocity curve of the objects, except in the innermost
region where they are too peaky. This is a common problem in cosmological
simulations (\citealp[e.g.][]{2003ApJ...591..499A, 2007MNRAS.374.1479G,
2010MNRAS.408..812S, 2011MNRAS.410.1391A}; and less severe in
\citealp{2011MNRAS.417..154S}), in which differences between the peak circular
velocity and the velocity at large radii are larger than in most observed
galaxies \citep{2006ApJ...640..751C}.

\begin{figure}
\includegraphics[width=1.0\columnwidth]{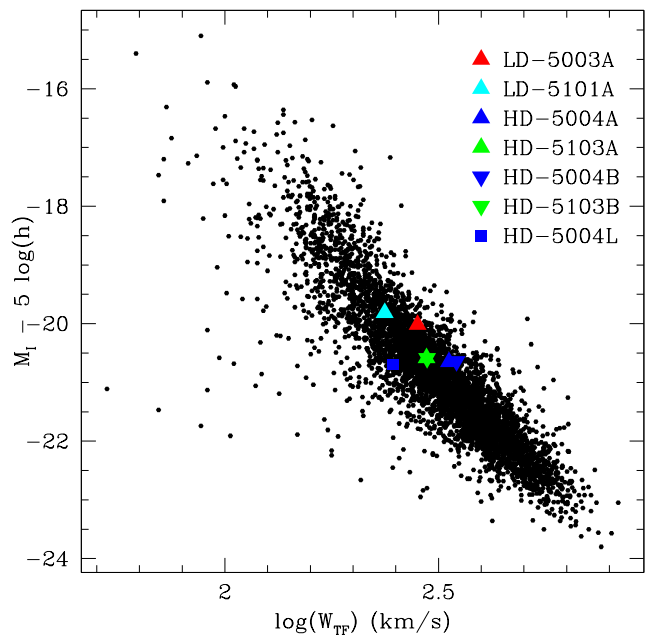}
\caption{Black dots correspond to the I-band Tully-Fisher relation from the
catalogue of \citet{2007ApJS..172..599S}. Coloured triangles represent our
simulated galaxies. The circular velocity corresponds to that measured at $R =
2.2 r_s$, and magnitudes have been computed using the
\citet{2003MNRAS.344.1000B} models.}
\label{tully}
\end{figure}

In Fig. \ref{tully} we show the I-band Tully-Fisher relation for our objects
(coloured triangles) as well as for the sample of spiral galaxies from the
catalogue of \citet{2007ApJS..172..599S}. All the simulated objects lie within
the region defined by the observational data scatter, although with some
systematic trend to be fainter than the average for spirals of similar rotation
speed.

\subsection{Angular momentum}

\begin{figure*}
\begin{tabular}{cc}
\includegraphics[width=1.0\columnwidth]{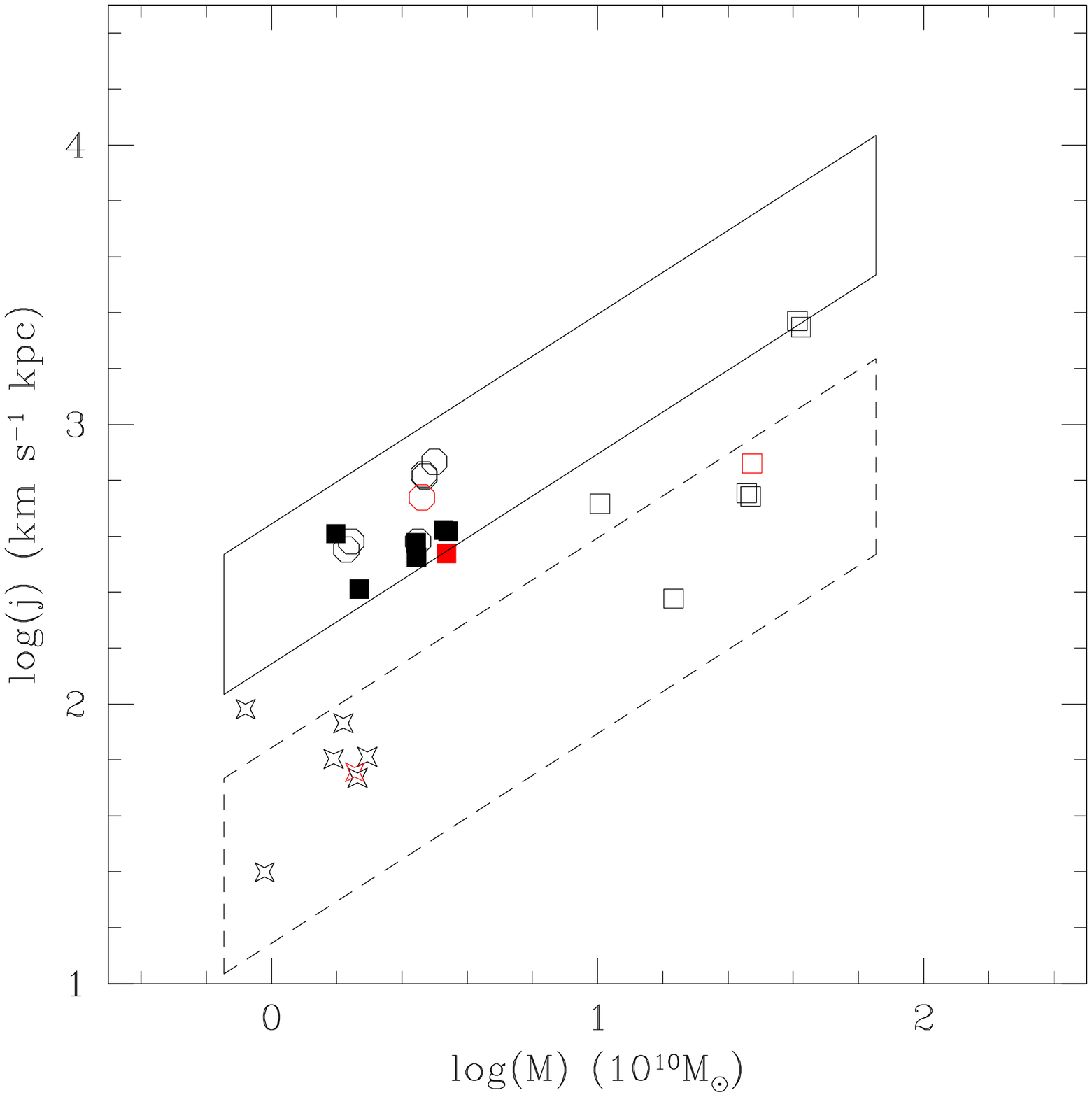} &
\includegraphics[width=1.0\columnwidth]{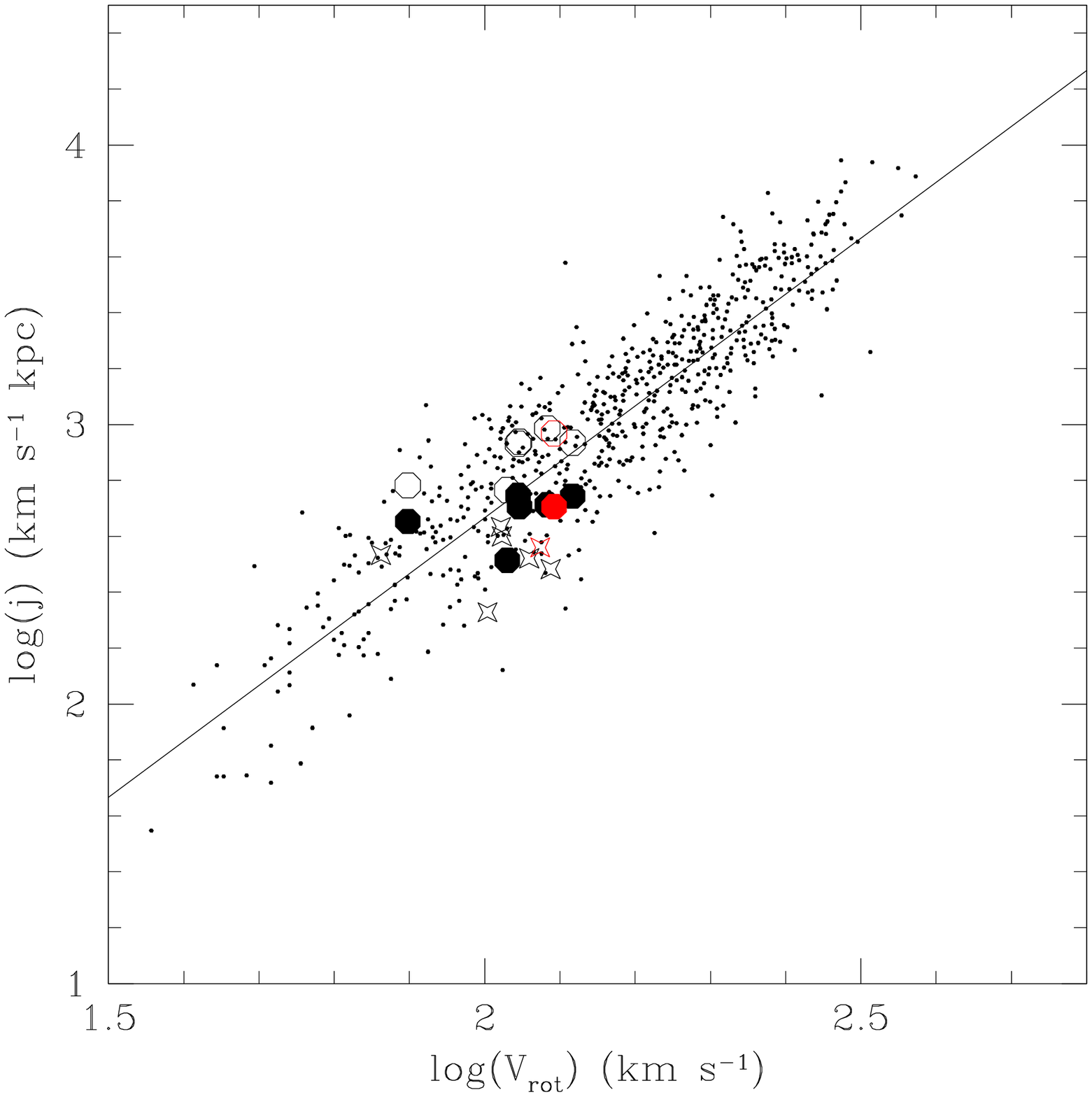}
\end{tabular}
\caption{Left panel: Specific angular momentum at $z=0$ versus the mass for
the dark matter haloes (open squares), baryonic discs (filled squares), and
stellar bulges (starred symbols). Open circles are computed using the same
methods than for observational galaxies: the specific angular momentum is
obtained from the rotational curve, and masses from the mass-luminosity
relation. Red symbols correspond to the low-resolution simulation. The limit
radial distances for this plot were $r_\mathrm{vir}$ for dark matter haloes,
$r_\mathrm{opt}=3.2r_s$ for discs and $r_e$ for bulges. The solid and dashed
boxes enclose the observational regions for spiral and elliptical galaxies,
respectively, as given by \citet{1983IAUS..100..391F}. Right panel: Specific
angular momentum versus rotation speed at $z=0$. Small black dots correspond to
late-type spirals compiled by \citet{1998astro.ph..7084N}. Filled circles
correspond to the simulated galaxies, assuming that the rotation velocity is
that measured at $r=2.2r_s$, and the angular momentum the sum of specific
angular momentum of baryonic particles. Open circles indicate the location of
the disc component, computed adopting the same angular momentum estimator as
applied to observed data, $j=2r_s v_c$. Starred open symbols represent angular
momentum and rotation velocity of the stellar component.}
\label{jmass}
\end{figure*}

In order to compare the angular momentum of the different components of our
simulated galaxies with observations, in Fig. \ref{jmass} we plot the specific
angular momentum ($j=J/M$) at $z=0$ as a function of mass ($M$) (left panel),
and as a function of the rotation speed (right panel). Recent works have made
use of similar representations \citep{2011arXiv1110.5635T,2011arXiv1112.0315S}
with the same purpose, since it provides a straightforward comparison with
observations and can help to pinpoint any angular momentum problems. In the
first plot, the specific angular momentum and mass of dark haloes (open
squares), stellar bulges (starred symbols) and baryonic discs (filled squares)
are represented separately. The mass and angular momentum of the dark matter
haloes have been estimated using particles within the virial radius, while
those of the bulges and discs have been estimated with stars\footnote{We will
refer to stellar particles in the simulation by the generic name `stars', even
though they are actually assumed to be single stellar populations of a given
mass (see Table \ref{table1}).} within $r<r_\mathrm{e}$ and baryons within
$r<r_\mathrm{opt}=3.2r_s$ (i.e., in a purely exponential disc $r_\mathrm{opt}$
is the radius enclosing 83 per cent of the disc mass), respectively
\citep{2003ApJ...597..878S,2003MNRAS.338..880S}. From this plot we can see
that the disc components and the dark haloes have comparable specific angular
momentum, so that disc particles have collapsed conserving, on average, their
angular momentum. On the contrary stellar bulges have been formed from material
that had lost most of its angular momentum, or that never acquired it. In the
figure we also include the boxes enclosing the observational regions covered by
spirals (solid line) and ellipticals (dashed line) as given by
\citet{1983IAUS..100..391F}.

Given that the observational estimates of specific angular momentum and mass
are derived from observational parameters, we also adopt an observational-like
approach, and compute the specific angular momentum and masses of our discs,
applying the method of \citet{1983IAUS..100..391F} directly to our simulated
discs (open circles in left panel of Fig. \ref{jmass}). The specific angular
momentum is obtained from the rotation curve as $j=2v_cr_s$ where
$v_c=V_{\mathrm{circ}}(R_{25})$ is the circular velocity at $r_{25}$, and
masses are derived assuming a constant mass-luminosity relation $(M/L_B)=3.0$.
We see from this figure that all simulated discs lie inside the region defined
by observed spirals and close to the values computed using observational
methods, although the latter show greater values for the specific angular
momentum. Similarly stellar bulges lie in the region defined by ellipticals,
except for some cases that still evidence the orbital angular momentum of
merging structures.

Another way to quantify any potential angular momentum problem is to represent
$j$ as a function of $v_{\mathrm{rot}}$. This approach has been followed
recently by \citet{2009MNRAS.400...43C} and \citet{2003ApJ...591..499A} among
others. \citet{1998astro.ph..7084N} provides a compilation of specific angular
momentum of observed discs derived using the estimator described above:
$j=2v_cr_s$. In right panel of Fig. \ref{jmass} we plot the observed data
(small dots), along with our objects using three different estimators for $j$
and $v_{\mathrm{rot}}$. Filled circles represent the sum of specific angular
momentum of baryonic particles, and their rotational velocity
$v_{\mathrm{rot}}$ is that measured at $r=2.2r_s$. Open circles are computed
using the observational estimator for $j$, as in right panel of Fig.
\ref{jmass}, and the same estimator for $v_{\mathrm{rot}}$. Open starred
symbols represent the specific angular momentum of star particles. Their
$v_{\mathrm{rot}}$ is derived using the rotation curve of stars only. While all
points lie within the observationally-defined region, it can be appreciated
that points derived using the actual baryonic angular momentum (filled circles)
have lower angular momentum than the observational-like estimates (open
circles). This behaviour was also observed in left panel of Fig. \ref{jmass},
and points to the fact that observational estimates, assuming a flat rotation
curve, can under-estimate or over-estimate the real value of the specific
angular momentum, depending on the trend that the rotation curve actually has.
The rotation curves of our simulated galaxies, decrease with increasing radius
owing to the relative importance of the bulge, as do most simulated discs in
the literature. Another effect is that stellar components generally have lower
angular momentum than the total baryonic component. Given that we take into
account all the star particles in this estimate, including an old bulge and
thick disc, this is expected, since these components are not fully supported by
rotation.

\subsection{Resolution effects}\label{sec:robustness}

In order to test for resolution effects in our simulations, we have run another
instance (HD-5004L) of the HD-5004A object, but with half the spatial
resolution and a mass resolution of about 8 times lower than in the
high-resolution run. The gravitational softening length was scaled to
$\epsilon_g = 800$ $h^{-1}$ pc, as well as the density threshold for SF, which
was set to $\rho_*$ = 3$\times10^{-26}$ g cm$^{-3}$ in order to roughly have
the same overall SF efficiency as in HD-5004A (i.e., an object stellar to
baryonic mass ratio of $\sim 85$ per cent). The main characteristics of this
low-resolution simulation, as well as some general properties of the resulting
object, are shown in the last row of Tables \ref{table1}, \ref{table2} and
\ref{tabledt}.

We can see from Table \ref{table1} that the resulting virial radius and virial
mass at $z=0$ are nearly the same as those of the high-resolution counterpart.
Both runs give also very close values for the stellar (a difference of $\sim 4$
per cent) and gas (a difference of $\sim 7$ per cent) mass inside the virial
radius.

The photometric decomposition (see Table \ref{table2}) also evidences similar
structural properties in the low and high resolution simulations. The
bulge-disc parameters of HD-5004L are almost identical to those of HD-5004A
(differences always smaller than $\sim 6$ per cent). The only exception is the
S\'ersic parameter $n$, which implies that the stellar distribution in the
innermost region is less concentrated in the low-resolution run. The D/T ratios
(see Table \ref{tabledt}) obtained both from photometric fits and a `kk-means'
decomposition are also rather similar to those measured in HD-5004A.

Kinematic properties do not appear significantly affected by resolution effects
either. Indeed, the specific angular momenta for the low-resolution dark matter
halo, stellar bulge and baryonic disc (red symbols in left panel of Fig.
\ref{jmass}) lie very close to their high resolution counterparts. Indeed, the
specific angular momenta for the low-resolution baryonic disc (red symbols in
right panel of Fig. \ref{jmass}) lie very close to their high resolution
counterparts. The circular velocity curves of HD-5004L and HD-5004A are also
quite similar, specially at large radii. As in other works
\citep[e.g.][]{2011MNRAS.417..154S}, the major differences are located in the
innermost regions, up to 6 kpc, and they are mainly due to the different
distribution of mass. In particular, the peak circular velocity diminishes
about 8 per cent in run HD-5004L. In this object the contribution of gas to the
total curve is negligible, so the difference between maximum values are mainly
due to the stellar distribution. Finally, in Fig. \ref{tully} we have also
included the Tully-Fisher relation for the low-resolution counterpart of object
HD-5004A, blue square dot. This object has a similar total magnitude but its
baryons are rotating slower, so the final low-resolution galaxy is brighter for
spirals of similar rotation speed.

In view of these results, we conclude that the low and high resolution objects
have converged in their structural and kinematic properties for the purpose of
our analysis.

\section{The fine structure}\label{sec:fineST}

\subsection{Dynamical decomposition method}
\label{sec:kk-means}

In order to characterise the properties of different components in the
simulated galaxies, we have devised a procedure to segregate star particles
into the main components, following the models proposed by
\citet{2003ApJ...597...21A} and \citet{2009MNRAS.396..696S}.

\begin{figure}
\begin{center}
\includegraphics[width=1.0\columnwidth]{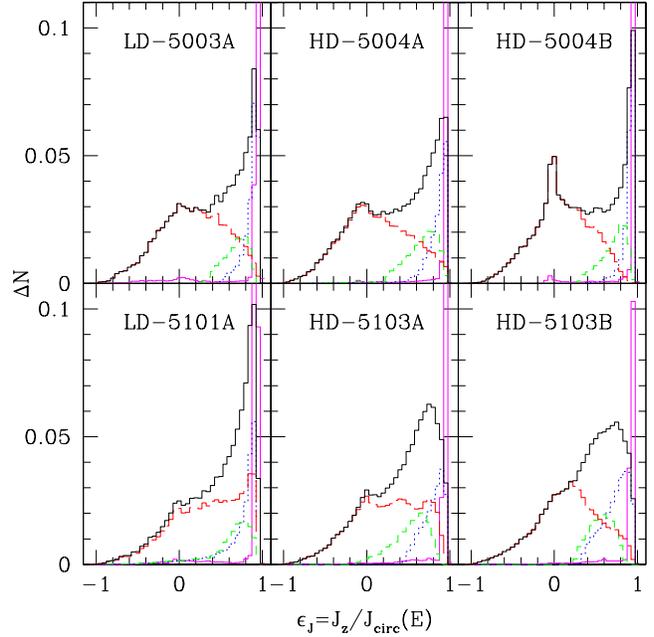}
\end{center}
\caption{Mass fraction of stars (black) and gas (magenta) for simulated
galaxies at $z=0$, as a function of $\epsilon_J = j_z/J_C$. Blue dotted, green
short dashed and red long dashed lines stand for mass fractions of stars in the
thin disc, thick disc and spheroid as decomposed by `kk-means' (see text for
details).}
\label{eps}
\end{figure}

The existence of a thin gaseous disc defines a symmetry axis, the $z$-axis,
determined by the angular momentum of gas particles in the inner regions of
haloes at $z=0$. We use this axis to orientate the object and then compute the
eccentricity parameter, $\epsilon_J$, as the ratio of the $z$-component of the
angular momentum of each baryonic particle, $j_z$, and the angular momentum
expected for a particle with the same binding energy in a circular orbit,
$J_C(E)$ \citep{2003ApJ...597...21A}. In Fig. \ref{eps} we plot the
distribution of this parameter, $\epsilon_J$, for stars (black line) and gas
(magenta line) in all simulated objects. As we would expect most gas particles
are distributed in circular orbits, with $\epsilon_{Jgas} \sim 1$. The
distribution of stars, on the contrary, typically shows two peaks: one at
$\epsilon_{J*}\sim1$ and a second one at $\epsilon_{J*} \sim 0$. This
distribution is quite similar to that previously found by other authors
\citep[e.g.][]{2003ApJ...597...21A, 2009MNRAS.396..696S}, and reveals the
presence of two main components: a cold disc in rotational support and a hot
spheroid supported by dispersion.

Although the distribution of $\epsilon_J$ allows us to infer the relative
importance of disc and spheroid, it is not enough for classifying individual
stars into different components. In order to assign individual stellar
particles to the different galactic components (thin and thick disc, bulge and
halo), more parameters are needed in addition to the eccentricity $\epsilon_J$.
In this work we will also use the total binding energy, $E$, computed using all
particles inside the virial radius, and the ratio of the projected angular
momentum of each baryonic particle on the plane of the disc, and the angular
momentum expected if the same particle was in a circular orbit $j_p/J_C$.

\begin{figure}
\begin{center}
\includegraphics[width=1.0\columnwidth]{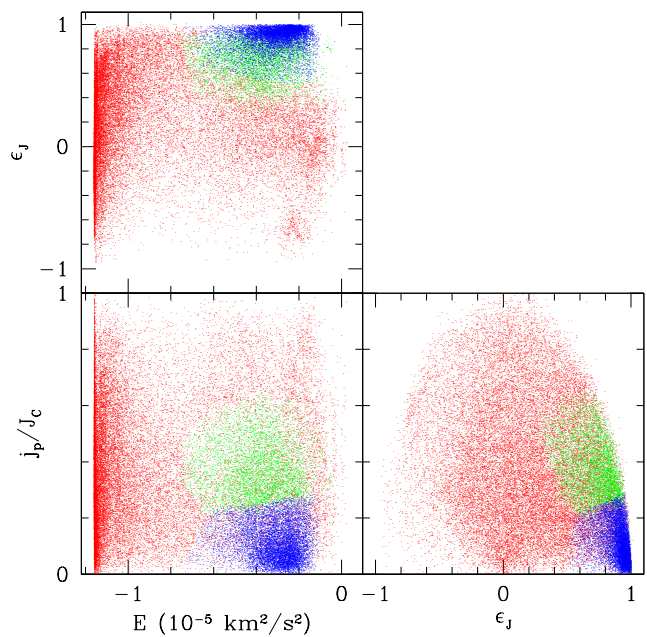}
\caption{Thick discs stars (green), thin disc stars (blue) and spheroid stars
(red) decomposed by `kk-means' in the 3D space defined by $\epsilon_J$
$j_P/J_C$ and $E$, the total binding energy computed using all particles inside
the virial radius, for simulated galaxy LD-5003A.}
\label{fig7}
\end{center}
\end{figure}

As it can be seen in Fig. \ref{fig7}, where we plot all the stars in object
LD-5003A, there exist heavily populated regions in the space defined by these
three parameters. Those regions can be easily identified as the spheroid and
the thin disc. In addition we also can recognize an intermediate, well
populated region, which is also separated by means of these parameters and that
we identify as the thick disc.

In order to assign stars in different components we use an unsupervised
clustering algorithm, the `kk-means' method \citep{Scholkopf1998,
Karatzoglou04kernlab, Dhillon04}, which is the weighted kernel version of the
`k-means' algorithm. The `kk-means' method uses the `kernel trick' in order to
implicitly project all data into a non-linear feature space using a Gaussian
kernel. Since the points are now located in a vector space of greater
dimensionality, it is able, in principle, to capture clusters that are not
linearly separable in input space, but are so in feature space. The algorithm
attempts to minimize the sum of the distance in feature space from points to
the assigned cluster centres, in the same way the standard k-means algorithm
does. It proceeds iteratively from an initial set\footnote{We run several
instances of the algorithm to ensure it properly converges to a stable set of
points that is independent of the initial choice of cluster centres.} of
cluster centres by assigning each point to the cluster with the closest centre.
Then, the centres are re-computed as the centroids of each cluster.

Given that the algorithm scales relatively poorly with the number of points and
cluster centres, we do not apply the algorithm to all stars in the main object.
Instead we select a random sample of points to classify according to this
method. The rest of stars are classified computing the Euclidean distance in
the feature space, and selecting the nearest neighbours decided by majority
vote. Finally we decide, studying the region spanned by each cluster, which
group corresponds to the different components of the object. We have performed
tests with different sample sizes to ensure the sampling procedure does not
affect the clustering results.

The unsupervised nature of this segregation procedure implies that no prior,
besides the expected number of components, have been assumed when separating
the different components. In its construction we have also avoided the use of
the properties (e.g., age or metallicity) expected for the stellar population
of each component. Therefore, this decomposition method will allow for an
unbiased study of many component properties and the connection with its history
of mergers and accretion events in the cosmological context.

\begin{figure*}
\includegraphics[width=1.0\textwidth]{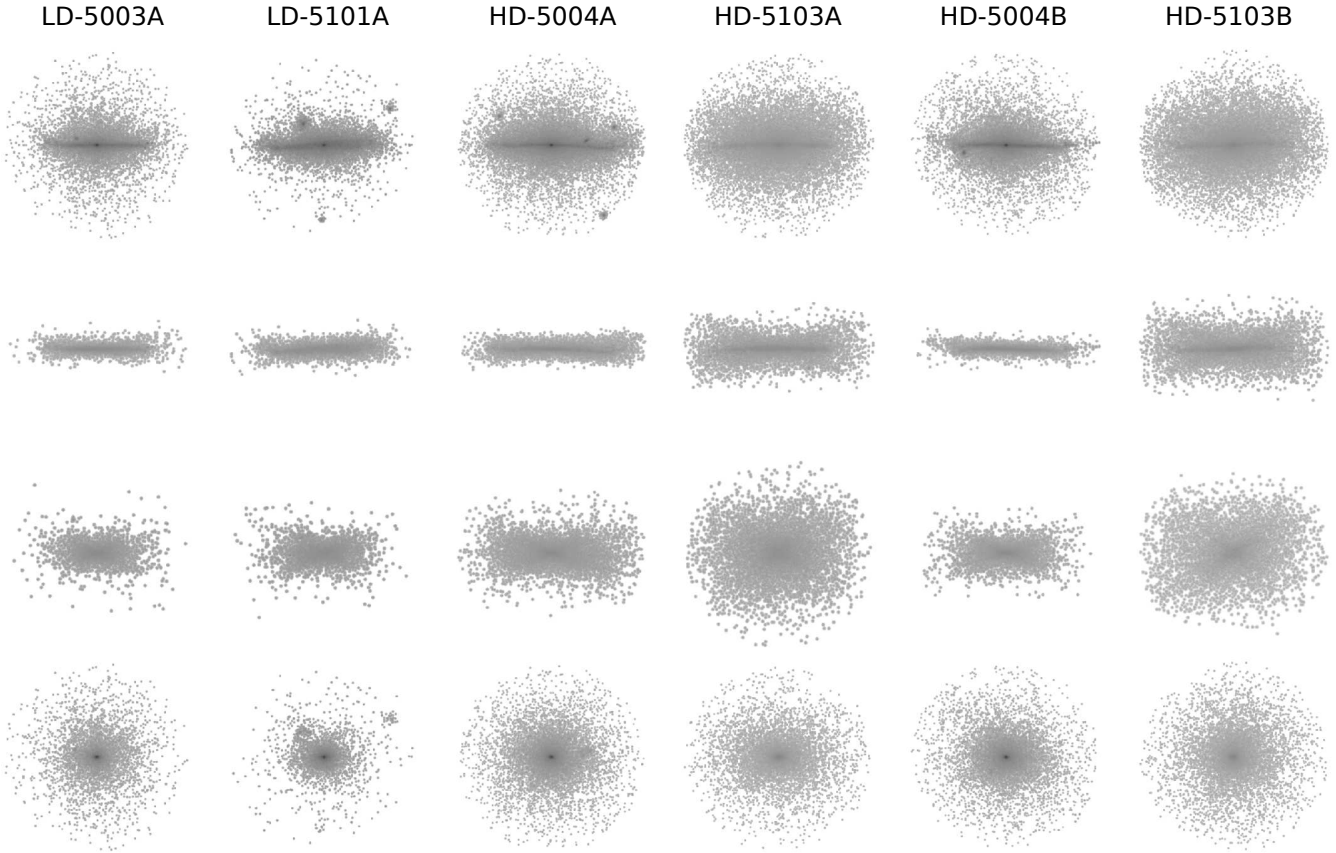}
\caption{Edge-on maps of stellar magnitudes ($r$-band) for the simulated
galaxies at $z=0$. All images are 40 kpc side. Top panels show the total
stellar content of each object, including satellites that will be excluded for
the analysis, and in next rows we draw, from second row to bottom, thin stellar
disc, thick stellar disc and spheroidal component as classified by our method.}
\label{fig8}
\end{figure*}

In Fig. \ref{fig8} we show the edge-on synthetic images of all stars in each
galaxy, including satellites that will be excluded for the subsequent analysis,
in top panels, and in subsequent rows we show the edge-on projection of stars
in different components: thin disc, thick disc, and spheroid, respectively. The
$\epsilon_J$ distributions of stars in each of these identified components are
shown in Fig. \ref{eps}, using the same colour code as in Fig. \ref{fig7}. In
this figure, we can appreciate that the spheroid component (in red), shows
almost no net rotation for most objects. Its distribution, however, tends to be
skewed towards positive values of $\epsilon_J$, this being specially clear for
object HD-5101A. We interpret this skewness as a manifestation of a possible
dual origin of the bulge, with a classical component supported by velocity
dispersion, and hence a symmetric $\epsilon_J$ distribution, and a pseudo-bulge
component with a net rotation (Obreja et al. submitted). In object HD-5103, A
and B, it is clearly seen that stars and gas peak at different $\epsilon_{J*}$
values. This may indicate that the stellar disc is not aligned with the gas
disc, probably due to the recent merger event in which it was involved.

\begin{table*}
\centering
\caption{Main properties of components: Stellar mass enclosed in the spheroid
(M$_\mathrm{sph}$), thin disc (M$_\mathrm{thin}$), and thick disc
(M$_\mathrm{thick}$). Fraction of mass in the thick disc compared to the total
stellar mass in the object (f$_\mathrm{thick}$). Scale lengths of the thin
(r$_{s1}$) and thick (r$_{s2}$) disc components, and the goodness of the fits
($\chi^2$). Scale heights of the thin (z$_{h1}$) and thick (z$_{h2}$) discs,
and the goodness of the fits ($\chi^2$). Finally, the asymmetric drift
(V$_\mathrm{lag}$).}
\begin{tabular}{l cccccccccccccc}
\hline
Galaxy & M$_\mathrm{sph}$ & M$_\mathrm{thin}$ & M$_\mathrm{thick}$ &
f$_\mathrm{thick}$ & r$_{s1}$ &$\chi^2$ & r$_{s2}$ & $\chi^2$ & z$_{h1}$ &
$\chi^2$ & z$_{h2}$ & $\chi^2$ & V$_\mathrm{lag}$\\
& 10$^{10}$ M$_{\odot}$ & 10$^9$M$_{\odot}$ & 10$^9$ M$_{\odot}$ & & kpc && kpc
&& kpc && kpc && km s$^{-1}$\\
(1) & (2) & (3) & (4) & (5) & (6) & (7) & (8) & (9) & (10) & (11) & (12) & (13)
& (14)\\
\hline
LD-5003A & 1.00 & 3.42 & 1.95 & 0.13 & 2.95 & 0.0022 & 2.87 & 0.0001 & 0.32 & 0.0291 & 1.49 & 0.0106 & -12.31\\
LD-5101A & 0.52 & 3.17 & 1.69 & 0.17 & 3.83 & 0.0032 & 3.45 & 0.0002 & 0.45 & 0.0017 & 1.35 & 0.0102 & -4.87\\
HD-5004A & 1.86 & 5.98 & 5.43 & 0.18 & 4.16 & 0.0011 & 3.81 & 0.0001 & 0.42 & 0.0108 & 2.33 & 0.0015 & -16.53\\
HD-5103A & 1.73 & 5.65 & 5.42 & 0.19 & 4.19 & 0.0016 & 4.26 & 0.0006 & 0.67 & 0.0061 & 8.32 & 0.0004 & -27.75\\
HD-5004B & 1.74 & 5.79 & 3.89 & 0.14 & 3.58 & 0.0027 & 3.39 & 0.0014 & 0.34 & 0.0059 & 2.17 & 0.0050 & -5.41\\
HD-5103B & 1.65 & 7.80 & 4.65 & 0.16 & 3.81 & 0.0002 & 4.76 & 0.0004 & 1.01 & 0.0005 & 13.4 & 0.0005 & -28.51\\
\hline
\end{tabular}
\label{table3}
\end{table*}

Table \ref{table3} summarises the main properties of each component. We see
from this table that the fraction of stars assigned to the thick disc is $\sim
16$ per cent of the total stellar mass, in agreement with observational results
\citep[e.g.][]{2006AJ....131..226Y, 2010arXiv1010.5276C}, in which the fraction
of stars locked in thick disc ranges from $\sim 10$ to $\sim 35$ per cent.
Columns (6) to (9) in this table, show the scale lengths of the thin (r$_{s1}$)
and the thick (r$_{s2}$) discs obtained by fitting their \textit{r}-band
luminosity profile with a simple exponential, and the goodness of those fits
($\chi^2$). From these quantities we see that both disc components have similar
radial scale lengths, as in observed galaxies \citep{2006AJ....131..226Y}.
Columns (10) to (13) show the scale heights z$_{h1}$ and z$_{h2}$ computed for
thin and thick disc by just considering the fraction of RGB stars in a solar
cylinder, as it will be explained in Section \ref{sec:heights}. We find that
for low-density objects the mean values of thin and thick disc scale heights
are lower than for objects in high-density regions, as we would expect. In any
case, the ratios of the scale heights for the thick and thin discs in our
simulated galaxies are found to be about 4 (excluding HD-5103), with a large
spread, as usually found for external galaxies
\citep[e.g.][]{2006AJ....131..226Y}, and in our Galaxy
\citep[e.g.][]{1983MNRAS.202.1025G, 2002ApJ...578..151S, 2008ApJ...673..864J}.
To end up, in the last column of Table \ref{table3} we display the asymmetric
drift (V$_\mathrm{lag}$), computed as the difference between mean rotational
velocity of thick and thin disc stars. These values show that the thick disc
rotates at a slower pace than the thin disc as it occurs in our Galaxy
\citep{2003A&A...398..141S} and in some external galaxies
\citep{2005ApJ...624..701Y}.

Finally, in Table \ref{tabledt} we show different D/T ratios measured with
different techniques. D/T$^\mathrm{P}$ represents values obtained from 1D fits
of the total i-band luminosity profile. The next column shows the
D/T$^\mathrm{P}_\mathrm{kk}$ ratios obtained when the luminosity profile of
particles in the `kk-means' spheroidal component is fitted with a S\'ersic
function and, separately, the luminosity profile of particles in the `kk-means'
disc components is fitted with a single exponential. Finally,
D/T$^\mathrm{L}_\mathrm{kk}$ gives the direct ratios, without any fit, of the
`kk-means' disc to total luminosity. We see from this table that the two fitted
ratios (D/T$^\mathrm{P}$ and D/T$^\mathrm{P}_\mathrm{kk}$) have roughly similar
values for each object, with a systematic tendency for
D/T$^\mathrm{P}_\mathrm{kk}$ to be slightly lower than D/T$^\mathrm{P}$. On the
contrary, the non-fitted direct ratios, D/T$^\mathrm{L}_\mathrm{kk}$, of the
`kk-means' components are significantly lower than D/T$^\mathrm{P}$. These
results are in agreement with those found in \citet{2010MNRAS.407L..41S}, where
our fitted and non-fitted D/T ratios must be compared to their photometric and
kinematic D/T ratios, respectively. As noted by those authors, the difference
obtained from these two methods can be partly explained from the fact that, in
the `kk-means' (or kinematic) decomposition, disc stars are not present or
hardly present within the inner regions, while the fitted photometric
decomposition assumes an exponential profile starting at the very centre of the
galaxy. These results show that, although the photometric D/T ratios can be
employed to compare the properties of simulated and real galaxies, kinematic
properties of galaxies are not properly inferred from such D/T ratios
\citep{2010MNRAS.407L..41S}.

\begin{table}
\centering
\caption{Disc to total ratios measured from different methods. See the text for
an explanation of the different methods used for the computation of these D/T
values.}
\begin{tabular}{l cccc}
\hline
Galaxy & D/T$^\mathrm{P}$ & D/T$^\mathrm{P}_\mathrm{kk}$ & D/T$^\mathrm{L}_\mathrm{kk}$\\
\hline
LD-5003A & 0.70 & 0.61 & 0.34\\
LD-5101A & 0.83 & 0.75 & 0.44\\
HD-5004A & 0.68 & 0.62 & 0.39\\
HD-5103A & 0.68 & 0.56 & 0.40\\
HD-5004B & 0.76 & 0.62 & 0.36\\
HD-5103B & 0.56 & 0.57 & 0.49\\
HD-5004L & 0.72 & 0.53 & 0.42\\
\hline
\end{tabular}
\label{tabledt}
\end{table}

\subsection{Observational-like samples}

A full decomposition method like that described in the preceding Section
\ref{sec:kk-means}, where each stellar particle is assigned to a galactic
component, is very useful for the modelling of the galactic structure and
formation processes, as well as for comparison with other simulation-based
studies. However, it can not be applied to observational data. In order to
facilitate the comparison of our results with those available from
observations, we have also selected spheroidal and disc subsamples based on two
different criteria similar to those frequently used in observational studies.
Each of these observational-like samples are not intended to contain all the
stellar particles of a given component, but just a subset of particles easily
identified in observations and that mostly belong to such a galactic component.

The first criterion is mainly founded on the positions of star particles in the
galactic object. The bulge sample consists of particles in the innermost region
($r \leq 5 r_e$) above and below the galactic plane ($2 z_{h1} \leq \arrowvert
z \arrowvert \leq 5 z_{h1}$) to minimize the contribution of disc stars. The
two disc samples are selected in a cylindrical shell $2 r_s \leq r \leq 3 r_s$,
which minimizes the contribution from bulges, and imposing a minimum rotation
speed of 50 km s$^{-1}$ in order to minimize the contribution from halo stars
\citep{2009MNRAS.400L..61S}. The distance to the galactic plane determines the
difference of both samples: stars located close to this plane ($ \arrowvert z
\arrowvert \leq 2 z_{h1}$) constitute the thin disc sample, while those at
farther heights ($ z_{h2} \leq \arrowvert z \arrowvert \leq 3z_{h2}$)
constitute the thick disc sample.

The second method uses kinematic information to select disc stellar samples, as
employed in Galactic stellar studies \citep{2003A&A...398..141S,
2003A&A...410..527B}. In those works, thin and thick disc stars are selected by
assuming that the Galactic space velocities of such stellar populations have
Gaussian distributions
\begin{equation}
f(U,V,W)\propto exp (-\frac{U^2}{2\sigma_U^2}-\frac{(V-\langle
V\rangle)^2}{2\sigma_V^2}-\frac{W^2}{2\sigma_W^2}),
\end{equation}
where U, V and W are the radial, azimuthal, and axial components of the stellar
velocity, respectively, and $\sigma_U$, $\sigma_V$ and $\sigma_W$ the velocity
dispersions found for thin and thick disc Galactic stars \citep[see
e.g.][]{2003A&A...398..141S}. Knowing these values, stars in the Galactic disc
are classified according to their major probability of belonging either to the
thin or to the thick disc. In our simulations, however, we can not use the
values of the Milky Way velocity dispersions, being necessary to compute them.
To do that, we first selected a sample of stars in `solar neighbourhood
volumes', equivalent to cylindrical shells normal to the disc plane between two
and three scale lengths of the thin disc ($2 r_s \leq r \leq 3 r_s$), which
will be mainly populated by disc stars, and then we segregated them in two
components according to their kinematics and metallicity, as in
\citet{2003A&A...398..141S}. These two new samples are employed to compute the
characteristic velocity dispersions that we will apply in the classification of
stars in the disc, not only in the `solar neighbourhood volume'. According to
\citet{2003A&A...410..527B}, in order to minimize the contribution of the thin
disc stars in the thick sample (and vice-versa), we classify as thick disc
stars those which are at least ten times more likely to be a thick disc star
than a thin disc star, i.e., those with $f_{TD}/f_{D} \geq 10$. Similarly, thin
disc stars are selected as those with $f_{TD}/f_{D} \leq 0.01$.

\section{Vertical density profiles}

\subsection{Scale heights of the disc components}\label{sec:heights}

The variation of stellar density with the distance \textit{z}, above or below
the Galactic plane, can be determined by counting faint stars in fields centred
on the Galactic poles. Stellar counts in the solar neighbourhood result in
vertical profiles which can be fitted with a double exponential function. This
fact was interpreted by \citet{1983MNRAS.202.1025G} as due to the presence of
two different disc components, thin and thick, embedded in the Galaxy. Values
measured for this double fit rises to $\sim350$ pc for thin disc scale height,
and ranges from $750$ to $1450$ pc for the thick one
\citep[e.g][]{1983MNRAS.202.1025G, 2002ApJ...578..151S, 2008ApJ...673..864J}.
In external galaxies, however, resolving stars is only possible for a small
number of galaxies, but the presence of thick discs have been photometrically
detected as an excess flux at large galactic latitudes
\citep{2002AJ....124.1328D, 2006AJ....131..226Y}. These works show that thick
discs are ubiquitous in disc galaxies with scale heights spanning a wide range
of values, but larger than the scale height of the embedded thin discs by a
factor $\sim 2$. \citet{2009MNRAS.395..126I} measured the scale height for the
Milky Way analogue NGC 891 from stellar counts and obtained a value, $z_h =
1.44 \pm 0.33$ kpc, quite similar to that observed in the Galaxy.

\begin{figure*}
\includegraphics[width=1.0\textwidth]{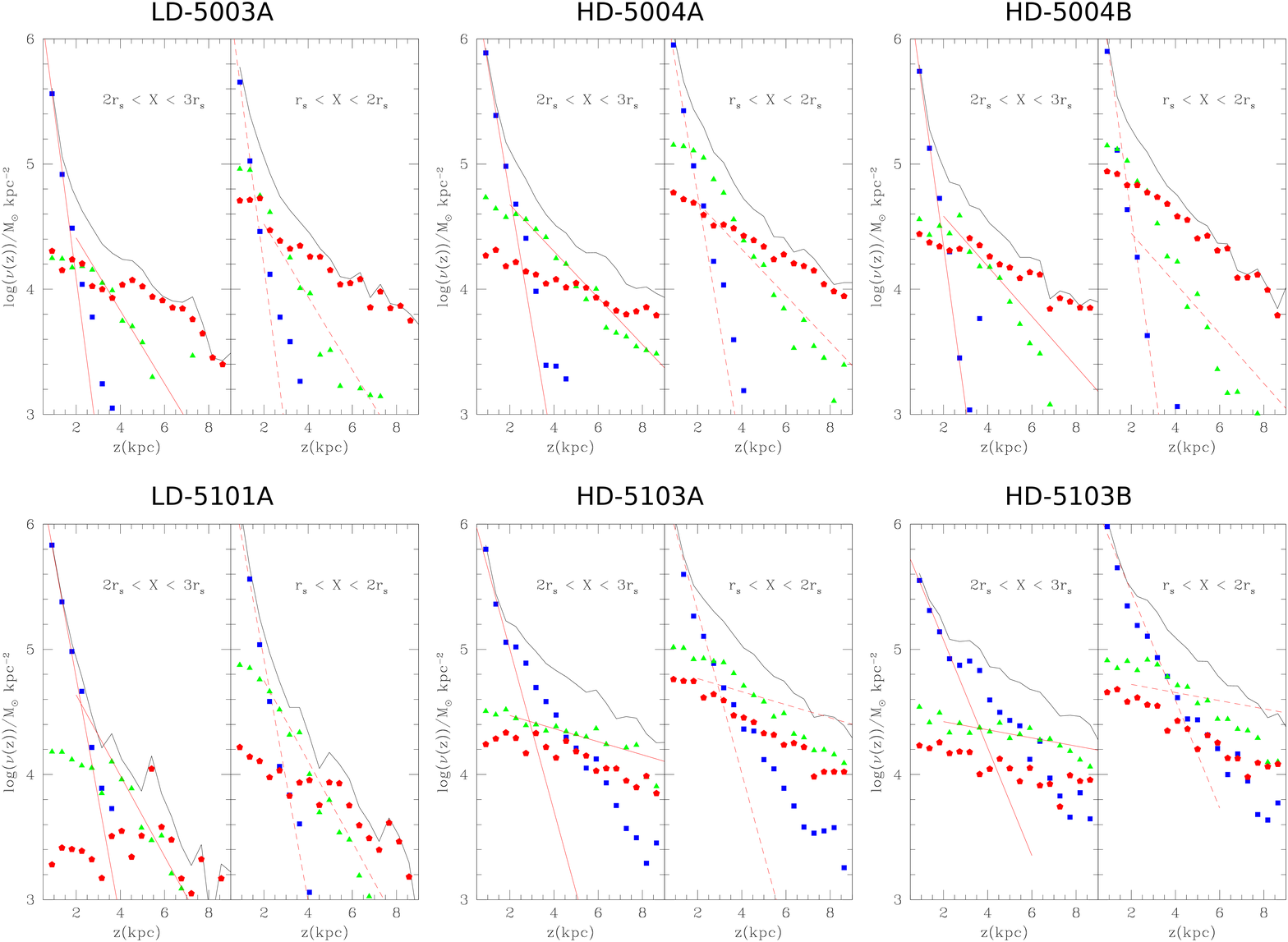}
\caption{ Vertical density profiles of all stars within two vertical cuts of
same width and different distance to the galactic centre. Black solid lines,
blue squares, green triangles and red pentagons represent the total, thin disc,
thick disc and spheroid stellar counts in each section, respectively. Red solid
lines at the left panels stand for the fits of thin and thick discs in the
outermost region. Red dashed lines at the right panels show an exponential fit
to the profiles in the internal cut constrained to have the same scale heights
for both, the thin and thick disc, as those respectively measured in the
outermost region.}
\label{vprof2}
\end{figure*}

In order to compute the vertical density profiles of simulated galaxies, we
have followed a procedure similar to that commonly used in observational
studies. The contribution of spheroidal components was minimized by just
considering stellar particles with projected galactocentric distances in the
range $2<r/r_s<3$. We then counted the number of RGB stars implied, for the
assumed IMF, by the age and metallicity of the stellar particles lying within
different vertical bins. The vertical profile of the surface density was then
obtained for each simulated galaxy by averaging several edge-on projections.
Such profiles are shown (black solid lines) in the left panels of Fig.
\ref{vprof2}, where we also represent the contribution of the thin (blue
squares) and thick (green triangles) discs and the spheroid (red pentagons)
decomposed by `kk-means'. The corresponding exponential fits of both disc
components are also shown (red lines) in Fig. \ref{vprof2}, and their values
are summarised in columns (10) and (12) of Table \ref{table3}. Fits of the thin
disc take into account only the innermost part of the profile, similarly the
thick disc fits ignore the innermost part, dominated by the thin disc.

The scale heights obtained in our low-density objects have mean values of
$\sim0.4$ kpc\footnote{Note that thin disc scale height is close to our
gravitational resolution, and thus it should be regarded with caution.} for the
thin disc and $\sim1.4$ kpc for the thick component. In the high-density
objects these scales are somewhat higher, specially those of the thick discs,
which have scale heights significantly larger than 2 kpc. This is likely due to
the more active history of mergers and accretions for galaxies in crowded
regions. Indeed, the object HD-5103A (and its counterpart HD-5103B with less SF
efficiency) has suffered a recent merger and for this reason it is perturbed at
$z=0$. As can be seen in the figure its vertical profile is almost exponential
and it is dominated by thick disc stars, even at farther heights where halo
stars are supposed to be the dominant population (red pentagons in Fig.
\ref{vprof2}). In any case, the ratios of the scale heights for the thick and
thin discs ($z_{h2}/z_{h1}$) in our simulated galaxies are found to be about 4,
with a large spread, as usually found for external early spirals \citep[see,
e.g.,][focused on Sd galaxies, but containing a compilation for S0-Sb
galaxies]{2006AJ....131..226Y}.

\subsection{Radial variation of the disc scale heights}

Another important feature of observed thick discs is that their vertical
profiles do not significantly depend on the galactocentric distance, i. e., the
scale height seems to be roughly constant with radius
\citep{2003A&A...409..523R,2009MNRAS.395..126I}. Such a result is difficult to
be reconciled with a merger-induced scenario for thick disc formation, which
inevitably leads to flared thick discs \citep{2009ApJ...707L...1B}, and favours
models where the disc thickening results from internal processes such as
gravitational instabilities.

In order to test whether the scale height changes with radius, we have followed
the procedure employed by \citet{2009MNRAS.395..126I} for an observed external
galaxy. We have drawn (right panels of Fig. \ref{vprof2}) the vertical density
profile of stars within a region ($1<r/r_s<2$) inner than in the preceding
analysis. For this internal region we have computed the exponential fits of
both discs components constrained to have the same scale heights, $z_{h1}$ and
$z_{h2}$, as in the external one. We find that, for the simulated objects
(LD-5003A and LD-5101A) in low-density regions of the Universe, roughly
constant $z_h$ values would result in acceptable fittings for both the inner
thin and thick. By contrast, for objects in high density-regions (HD-5004A and
HD-5103A, and their counterparts HD-5004B and HD-5103B with less SF
efficiency), roughly constant $z_h$ values give acceptable fits only for the
thin disc of some simulated galaxies, but clearly fails for the thick disc.
Therefore, for these latter objects we can not reject the merger-induced
scenario as a key ingredient for thick disc formation.

\begin{figure*}
\centering
\includegraphics[width=0.8\textwidth]{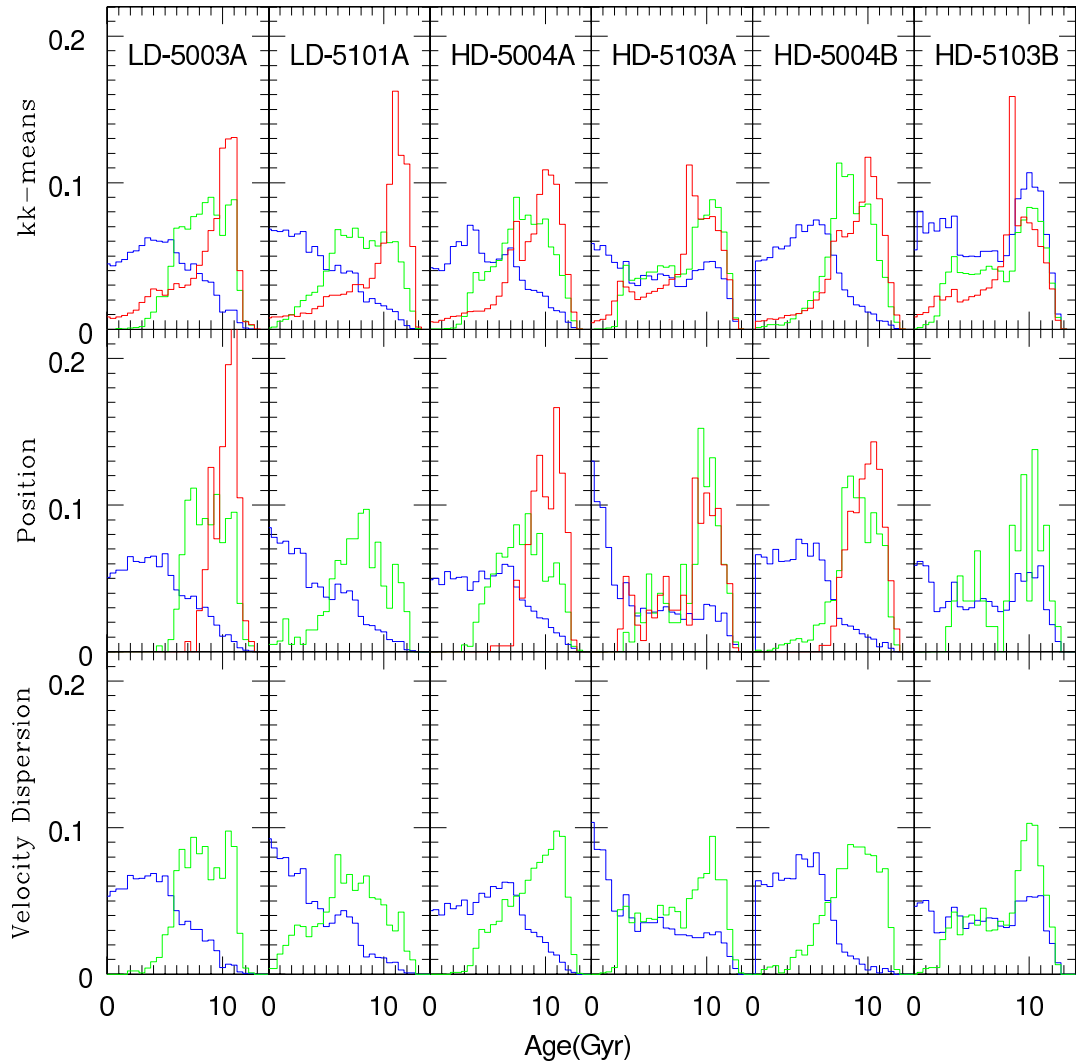}
\caption{Stellar age distribution normalised to the total mass of each
individual component. In each row we depict the age distributions obtained
with different classification methods, as indicated in the \textit{y}-axis. In
the first row we show age distributions for star particles in each component
identified by our classification method. In the second one samples are
selected by positions in the simulated galaxies, and in the last one, kinematic
criteria have been employed to identify stars in thin and thick disc. In all
panels, blue line refers to thin disc star particles, green is for thick disc
and red line stands for the spheroid.}
\label{fig10}
\end{figure*}

\section{Stellar ages}
\subsection{Age distributions of the different components}

\begin{figure}
\centering
\includegraphics[width=1.0\columnwidth]{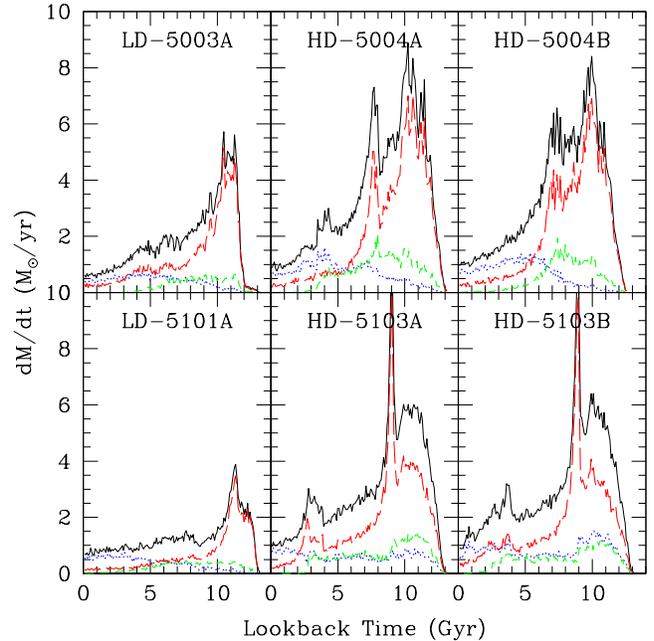}
\caption{
Star formation history for each component. Black solid line stands for the
total star formation rate in the object. Thin disc is represented with a blue
dotted line, and green short dashed line for the thick disc. Star formation
rate in the spheroidal component is plotted with a red long dashed line.
}
\label{sfr}
\end{figure}

Fig. \ref{fig10} shows the distribution of stellar ages for the different
components of our simulated galaxies. Each distribution was normalised to the
total stellar mass of the corresponding individual component. The different
rows in this figure depict the age distribution of samples selected from the
different criteria described in Section \ref{sec:fineST}: the first row refers
to `kk-means' decompositions, the second row to observational-like samples
built from position criteria, and the third row to observational-like samples
built from kinematic criteria. The corresponding median values of stellar ages
are summarised in Table \ref{table4}, while the star formation histories for
each object and its `kk-means' components are shown in Fig. \ref{sfr}. Note
that, for runs LD-5101A and HD-5103B, the observational-like samples for the
bulge component were too small for statistics and, hence, they are not depicted
in Fig. \ref{fig10}.

A clear trend is found in all simulated galaxies for any kind of sample. The
thin disc is always the youngest component with median ages ranging from 3.8 to
6.7 Gyr. Just a small fraction of the thin disc stars have ages older than 10
Gyr, while just a small fraction of stars in the spheroid are younger than 5
Gyr. The thick disc is instead populated by stars with intermediate ages. These
results are consistent with the observational estimates for the different
components in the Milky Way and nearby galaxies. Indeed, several
\citep[e.g.][]{1998A&A...338..161F, 2003A&A...410..527B, 2005A&A...433..185B,
2006MNRAS.367.1329R} agree that stars in the thick disc of the Galaxy have mean
ages greater than $8$ Gyr meanwhile the estimates for the thin disc stars lay
between $5$ to $7$ Gyr. \citet{2005A&A...433..185B} derived a mean age of $4.3
\pm 2.6$ Gyr for stars in the thin disc of the Galaxy, and $9.7 \pm 3.1$ Gyr
for a sample of thick disc stars. Similarly, for nearby galaxies, photometric
studies find that the off-plane regions are dominated by old stars ($>6$ Gyr)
\citep{2006AJ....131..226Y, 2002AJ....124.1328D}. In their study of the
integrated spectra of six galaxies, \citet{2008ApJ...683..707Y} obtained a
median age of $4.6$ Gyr for the thin disc, and ages ranging from $3.8$ to
$10.9$ Gyr (with a median value of $7.2$ Gyr) for the thick disc.
\citet{2011MNRAS.417..154S} have also found similar results for the study of
stellar populations in their simulated galaxies. The agreement between results
obtained with different simulation techniques and decomposition methods, could
be interpreted as a general feature in the age of stellar populations of
simulated galaxies.

\begin{table}
\centering
\caption{Stellar median ages in Gyr for different components of each simulated
galaxy. The three rows for each object correspond to a `kk-means'
decomposition, samples from position criteria, and samples from kinematic
criteria, respectively.}
\begin{tabular}{cccc}
\hline
Galaxy & Age$_{\rm sph}$ & Age$_{\rm thick}$ & Age$_{\rm thin}$\\
\hline
LD-5003A & $9.6\pm2.2$ & $8.7\pm2.3$ & $4.7\pm3.3$\\
      & $10.7\pm1.0$ & $9.0\pm2.0$ & $4.3\pm3.1$\\
      &              & $8.5\pm2.3$ & $4.1\pm2.9$\\
LD-5101A & $10.8\pm1.9$ & $8.1\pm3.0$ & $3.8\pm3.3$\\
      &              & $8.1\pm2.4$ & $3.3\pm3.2$\\
      &              & $7.2\pm2.9$ & $3.1\pm3.0$\\
HD-5004A & $9.7\pm2.2$ & $8.3\pm2.5$ & $4.7\pm3.3$\\
      & $10.3\pm1.6$ & $8.3\pm2.4$ & $5.0\pm3.6$\\
      &              & $9.5\pm2.5$ & $5.3\pm3.4$\\
HD-5103A & $9.0\pm2.9$ & $9.0\pm2.8$ & $5.7\pm3.8$\\
      &  $9.4\pm2.7$ & $9.8\pm2.3$ & $3.1\pm3.8$\\
      &              & $8.6\pm2.9$ & $3.8\pm3.6$\\
HD-5004B & $9.5\pm2.2$ & $8.6\pm1.9$ & $4.9\pm2.7$\\
      & $10.2\pm1.5$ & $9.3\pm1.8$ & $4.3\pm2.7$\\
      &              & $9.1\pm2.2$ & $4.3\pm2.6$\\
HD-5103B & $8.8\pm2.0$ & $8.7\pm2.9$ & $6.7\pm3.7$\\
      &  $9.9\pm1.8$ & $9.4\pm2.7$ & $6.2\pm3.8$\\
      &              & $9.2\pm2.8$ & $6.4\pm3.7$\\
\hline
\end{tabular}
\label{table4}
\end{table}

Focusing on star formation histories, we can see in Fig. \ref{sfr} that the
thin discs of all simulated galaxies are currently forming stars with a star
formation rate lower than about one solar mass per year. For almost all
simulations the bulk of stellar mass formed 10 Gyr before the present epoch,
populating preferably the spheroid. At low redshift, however, there is just
some residual star formation in this component. On the contrary, the thick disc
is not forming new stars at low redshifts. The stellar particles settled in
this component were formed during an extended epoch in the past, with a low
rate similar to that observed in the thin disc. In the last 10 Gyr merger and
accretion events triggered the star formation leaving a trace as peaks in the
star formation history. This is clearly visible in simulation HD-5103A and its
counterpart HD-5103B, which have been involved in a recent major merger. The
new stars formed during this event mainly increase the stellar content in the
spheroid and the thin disc. Differences in the SF parameters produce
differences in the response of simulated galaxies to merger events. In those
objects in which SF has been difficult, the SF occurs more quietly and during a
more extended epoch than in their counterparts with a more efficient SF, in
which those events produce sharper peaks.

\subsection{Age and colour gradients}

\begin{figure*}
\includegraphics[width=0.8\textwidth]{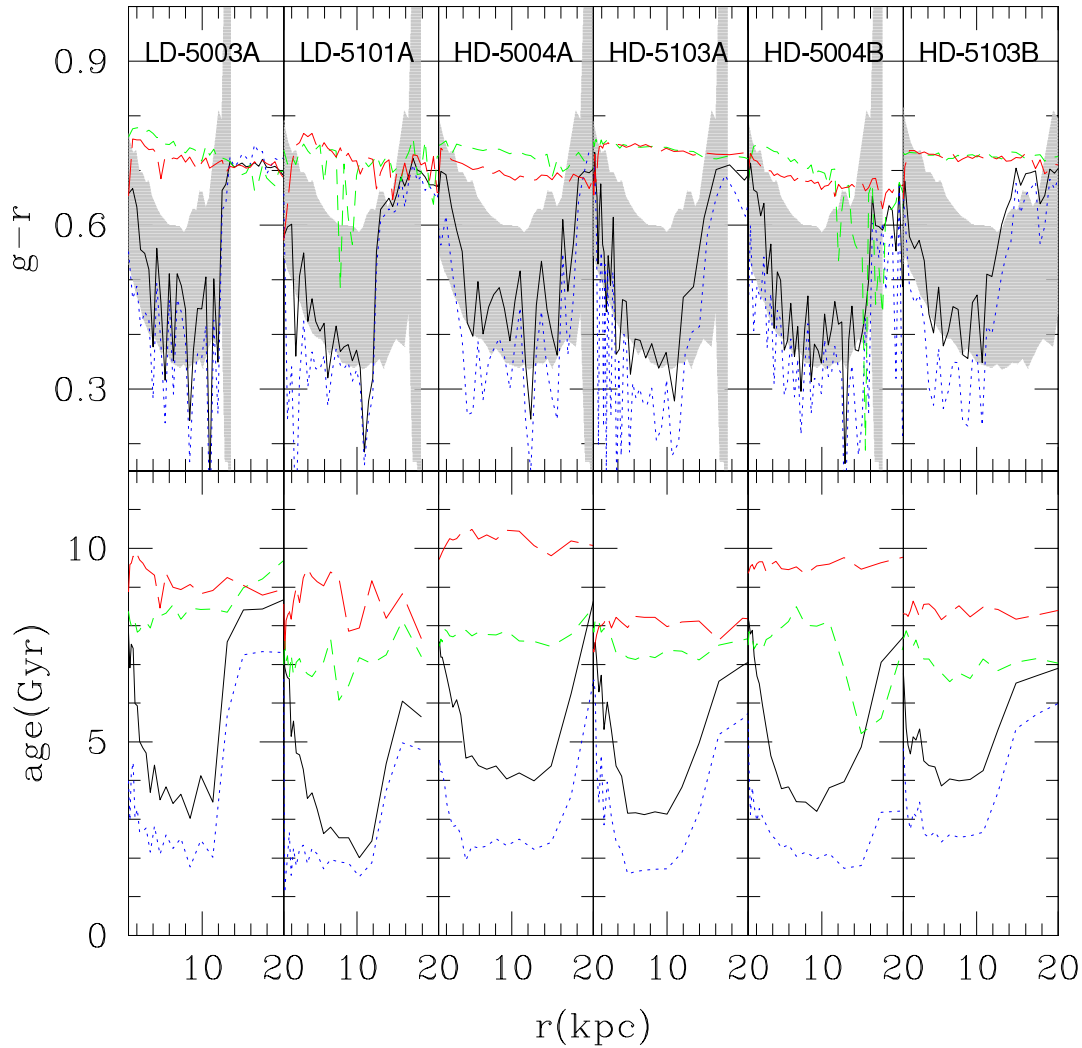}
\caption{
Top panels: Colour profiles for each component in the galaxy as seen face on.
As in previous figures solid line stands for the colours of all stars in the
galaxy. Thin disc stars are plotted using a blue dotted line, and the colour
gradient of thick disc stars is shown with a green short dashed line. Red long
dashed line stands for colour gradients in spheroid stars. For reference, the
observed colour profiles of 39 Type-II galaxies, as reported by
\citet{2008ApJ...683L.103B} are shown as a shaded grey area. Note that the
population synthesis used in this paper does not include dust effects. Bottom
panels: Mean stellar ages weighted by $r$-band luminosity for each simulated
galaxy. Line code is as in top panel and previous figure. Similar profiles
weighted by mass can be found in our previous paper
\citep{2009ApJ...705L.133M}.}
\label{color_age}
\end{figure*}

Colour profiles are related with galactic properties such as metallicity,
stellar age or dust content, and for this reason they have long been used for
exploring galactic properties in large galaxy samples and catalogues
\citep[e.g.][]{2010MNRAS.tmp.1828G}. Fig. \ref{color_age} shows the
\textit{g-r} colour and age profiles for each simulated galaxy. Black solid
lines depict the total profiles, while coloured dashed lines show the
individual profiles of each galactic component identified by the `kk-means'
method. From this figure we see that the colour profiles are dominated by the
thin disc contribution, which is responsible for their U shape: every galaxy
and thin disc in our sample is redder in the innermost region and becomes bluer
as we move away in the radial direction until a minimal value is reached. Then,
the stellar content becomes progressively redder. These results are in good
agreement with observations of Type-II spiral galaxies
\citep[e.g.][]{2008ApJ...683L.103B, 2008ApJ...683..707Y, 2010MNRAS.tmp.1828G},
where such U-shaped profiles are also found, with median colour gradients close
to zero. These features have been recently studied using cosmological
hydrodynamic simulations \citep{2009ApJ...705L.133M, 2009MNRAS.398..591S} and
tend to favour an inside-out formation scenario for the galactic discs.
According to such a scenario, star formation has moved progressively towards
outer regions of galactic discs. As a consequence, older and more metal-rich
stellar populations are located in the innermost regions of discs leading
negative colour gradients. Due to the secular evolution, stars migrate from the
inner to the outer parts of the galaxy, reaching the outskirts where the
specific star formation is very small. These outskirts progressively become
populated by old and red stars \citep{2009ApJ...705L.133M,
2009MNRAS.398..591S}.

In contrast to that found for the thin disc, the colour and age profiles
depicted in Fig. \ref{color_age} for the thick disc and spheroidal components
are quite uniform across the galaxy, with little sign of radial gradient. These
profiles are a clear indication of a different formation scenario for such
components: their stellar content is older and was formed in a shorter time
than in the thin disc (see also Fig. \ref{fig10} of age distributions). As in
observed galaxies \citep[]{2009MNRAS.395...28M}, age gradients in the spheroid
are almost zero, with just small deviations from a flat profile both towards
negative or positive values. These deviations are probably more visible for
objects in low density regions, where the spheroid shows a slightly negative
age gradient.

\section{Chemical properties}\label{sec:chem}

\subsection{Metallicity Distribution Functions}

Element abundances of individual stars and gas provide a valuable record of the
star formation and enrichment history during the early stages of galaxy
formation and evolution, and it has long been used to propose or test the
different mechanisms involved in such processes
\citep[e.g.][]{1995AJ....109.1095G, 2006MNRAS.367.1329R, 2006A&A...445..939S}.
The fact that kinematic and chemical properties of thick disc stars differ form
those of thin disc stars has been interpreted as an evidence of their distinct
nature, making possible the differentiation of stars from both components using
their chemical properties \citep[e.g.][]{2006A&A...445..939S,
2009MNRAS.399.1145S}. Otherwise \citet{2008A&A...484L..21M} found no
significant abundance differences between the bulge and the local thick disc
stars in their study, which is interpreted as a consequence of the rapid star
formation history of both components, rather than any deeper relation between
them.

In the Milky Way, thick disc stars have been found to be old and more
metal-poor and also $\alpha$-enhanced compared to stars in thin disc
\citep[e.g.][]{1998A&A...338..161F, 2003A&A...410..527B, 2005A&A...433..185B,
2006MNRAS.367.1329R, 2010ApJ...721L..92R}. To constrain the chemical evolution
of the Galaxy, samples of long-lived stars, as F and G dwarfs, have been
employed \citep[e.g.][]{1995AJ....109.1095G, 1998A&A...338..161F,
2003A&A...410..527B, 2005A&A...433..185B, 2006MNRAS.367.1329R}. The metal
distribution of thick disc stars is roughly characterized as a Gaussian with
mean iron abundance $\sim$ -0.7 to -0.5 dex \citep[e.g.][]{1995AJ....109.1095G,
2006A&A...445..939S}, with most stars between -0.5 to -1.0 dex, but with a long
tail of metal-poor stars which extends to about -2.2 dex, values usually
associated with halo stars. This metal-poor tail is interpreted by some authors
as an evidence of the complex structure within the thick disc
\citep[e.g.][]{2006A&A...445..939S,2010ApJ...712..692C}.
\citet{2008ApJ...683..707Y} used Lick metallicity-sensitive indices to derive
metallicity properties of thin and thick disc components in six external
galaxies. They found metal-poor and no significant $\alpha$-enhanced thick
discs, probably due to the insensitivity of Lick indices to the differences
found at low metallicities. The median metallicity value measured in the thick
and thin disc of these galaxies is [Z/H]=-0.6, as expected for low-mass
galaxies.

\begin{figure*}
\begin{center}
\includegraphics[width=0.8\textwidth]{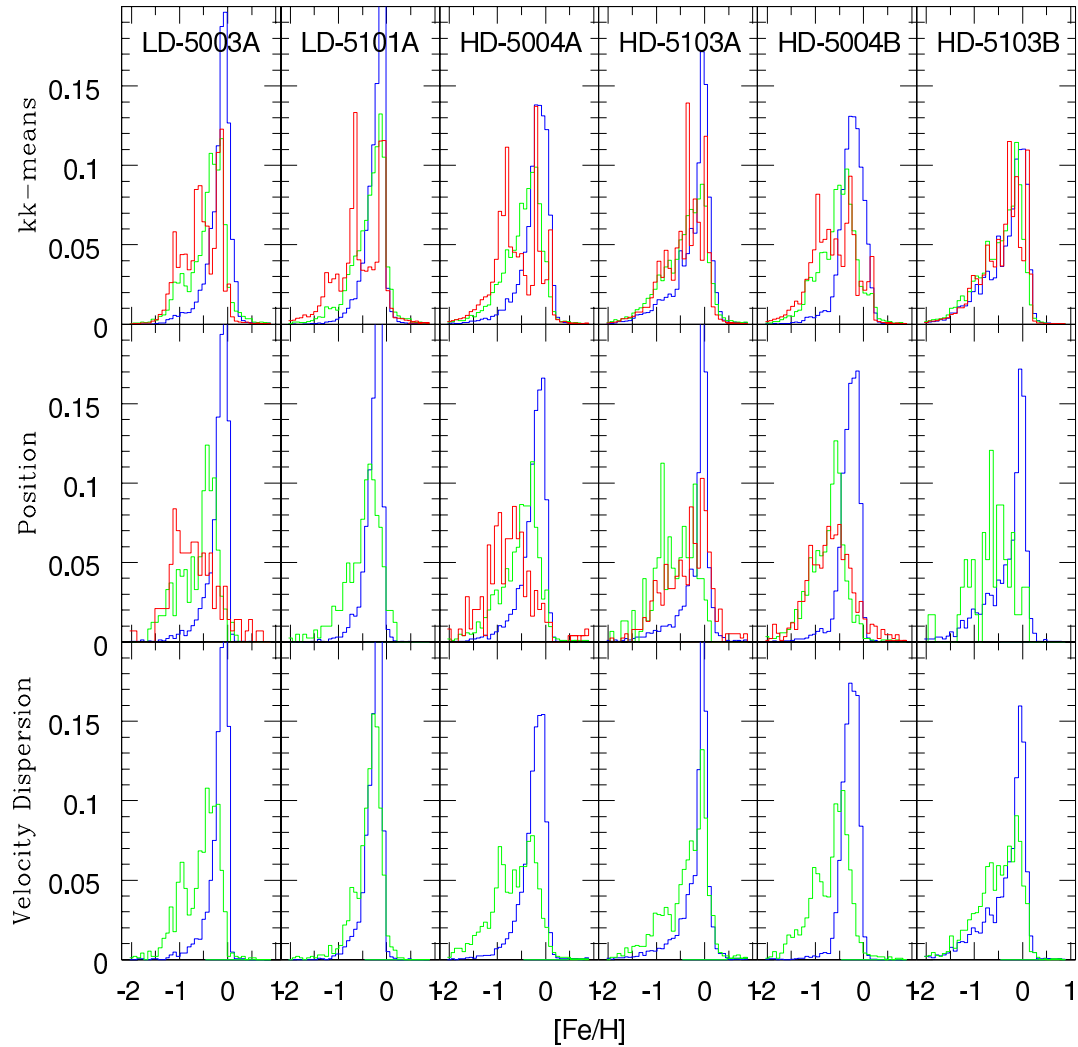}
\end{center}
\caption{Metallicity distribution function of stars in each galaxy component. As in Fig.
\ref{fig10} in each row is depicted the MDF of samples selected from the
different described criteria. Each component is represented by different
colours, as in previous figures.}
\label{mdf}
\end{figure*}

In Fig. \ref{mdf} we depict the MDF of each component in our simulated
galaxies\footnote{The solar abundances of \citet{1998SSRv...85..161G}, with
$Z_{\odot}=0.0169$ have been assumed for the subsequent analysis.}. As in Fig.
\ref{fig10} each row corresponds to samples selected from the different
criteria described in Section \ref{sec:fineST}, and each distribution is
normalised to the total stellar mass of the corresponding individual component.
In general we find that the thin disc is the most metal-rich component, with
median values close to the solar value, between -0.19 to -0.06. Moreover, the
distribution is quite similar for the different thin disc samples. The
spheroid, on the contrary, is the most metal-poor distribution, with median
values covering a wide range between -0.61 to -0.27. In this component a clear
difference is observed among the different samples. In those based on the
`kk-means' method we find a bimodal MDF, while such a bimodality is not
observed in the samples selected from position criteria, which exclude stars
near the galactic plane. Since the distribution of parameters $\epsilon_J$ and
$j_p$ for these high-metallicity stars in the spheroid clearly corresponds to
this component, such a high-metallicity peak seems to be a real feature of the
spheroid MDF, and not an artefact due to disc stars that eventually could be
misclassified as belonging to the spheroidal component. The bimodal MDF for
the spheroid seems more likely due to the different formation mechanisms of the
bulge of spiral galaxies: the low-metallicity peak would correspond to the
classical bulge, while the high-metallicity peak would correspond to the bulge
fraction formed from instabilities of the inner disc (Obreja et al. submitted).

With the only exception of simulation HD-5103A and its counterpart HD-5103B, in
which the stellar content is well mixed, the thick disc MDF shows an
intermediate behaviour between the thin disc and the spheroid. This component
has in fact a metal content poorer than in the thin disc, while richer than in
the spheroid. Their MDFs peak at values from -0.45 to -0.24, but their shape
is not as sharp as for the thin disc and extends to very metal-poor regions. In
the case of observational-like samples selected from kinematic criteria (third
row in Fig. \ref{mdf}) the thick disc MDF shows a clear bimodality, which could
be related with the metal-poor thick disc defined by some authors
\citep[]{2006A&A...445..939S, 2010ApJ...712..692C}.

\subsection{Metallicity variation along the disc}

The distribution of element abundances within galaxies is another important
feature in the comprehension of galaxy formation and evolution. First measures,
based in emission lines in large samples of HII regions in spiral galaxies
\citep[e.g.][]{1983MNRAS.204...53S, 1994ApJ...420...87Z, 1997ApJ...489...63G,
1998AJ....116.2805V}, found radial abundance gradients in almost all large
spiral galaxies, with inner HII regions richer than outer ones.
\citet{1983MNRAS.204...53S} measured an oxygen abundance gradient of
$\sim-0.07$ dex kpc$^{-1}$ in the Milky Way, a value quite similar to that
obtained by \citet{2006AJ....132..902L} fitting iron abundances of young stars
in the Galaxy, or \citet{2010A&A...511A..56P} using open clusters. This result
suggests that the chemical evolution of the Galactic disc is determined mainly
by local chemical evolution, because mergers would have produced more complex
abundance distributions. In external regions, however, steeper gradients are
found, with metal-poor stars at high radial distances
\citep[e.g.][]{2005AJ....130..597Y}. The existence of such a radial profile can
indicate that the unobserved inner stellar thin disc may be significantly more
metallic than the solar neighbourhood.

Observations of the Milky Way reveal that the different galactic components
have characteristic abundance patterns. \citet{2011MNRAS.412.1203N} show
distinctive abundances for the thin and the thick disc in the solar
neighbourhood, with the thick disc being significantly less enriched. The
spheroid is known to have a gradient with the outer (halo) parts being very
metal poor \citep{2008ApJ...684..287I}, and the inner bulge having
metallicities around the solar value \citep{2008A&A...486..177Z}.

\begin{figure*}
\begin{center}
\includegraphics[width=1.0\textwidth]{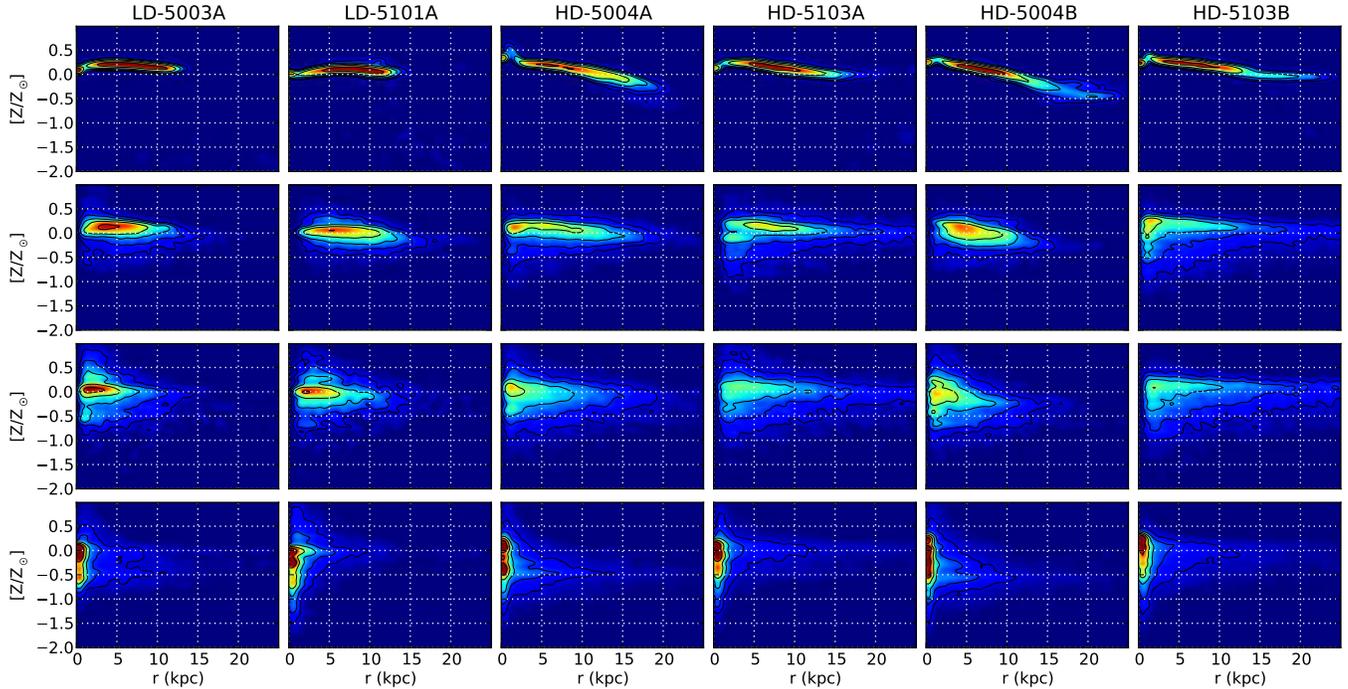}
\end{center}
\caption{Metallicity radial distributions for gas, thin disc, thick disc and
spheroid (from top to bottom).}
\label{metgr}
\end{figure*}

In Fig. \ref{metgr} we depict the metallicity vs. radius distributions for each
component identified by the `kk-means' method in our simulated galaxies. In
addition, the first row represents the gas component, with computed gradients
ranging from $-0.060$ dex kpc$^{-1}$ in the object HD-5004A to $-0.039$ dex
kpc$^{-1}$ in LD-5101A which is the less enriched galaxy.

At first glance, one can appreciate that the more `disky' a component is, the
lower its dispersion in metallicity for any given radius, with the spheroids
showing the highest range of variation in metallicity, and the thin discs the
lowest. The thick discs appear here as an intermediate component, with almost
the same radial extent as the thin disc but a higher dispersion in their
metallicity values. The least massive objects (LD-5003A and LD-5101A) show the
lowest metallicity gradient for the thin disc component, whereas a more massive
galaxy (HD-5004A and its counterpart HD-5004B) show a larger gradient, with the
run having the least effective set of SF parameters (HD-5004B) having a larger
gradient.

Given that our sample is very small, no conclusions can be drawn with respect
to the relation between morphological type (or D/T ratio) and metallicity
gradient, although there is a similar effect known from observations
\citep{1993MNRAS.261L..17E}. The difference between HD-5004A and HD-5004B,
given that it is the same object with a different set of SF parameters, is
however more telling. A reduction in the overall SF efficiency leads to a
steeper gradient. Given that the metallicity values in the inner part of both
objects is similar, the difference must be due to a reduction in the SF
efficiency in the outer parts of the disc, where the density threshold
criterion is not always met by the HD-5004B simulation. This effect is present
in observations of nearby spirals \citep{2008AJ....136.2782L,
2010AJ....140.1194B}, where a decreasing SF efficiency with density (or
galactocentric radius) is found, leading to extremely inefficient SF in the
outskirts of discs.

The spheroids show a rather complex structure in Fig. \ref{metgr}. For most
objects several bumps in the MDF of the bulge are apparent (see also Fig.
\ref{mdf}). As was explained in the preceding subsection, this is likely the
outcome of two different processes driving the formation of the bulge, with the
highest metallicity peak being associated with a pseudo-bulge and the lower
metallicity distribution being associated with the classical bulge. Further
results on the bulge abundances and their formation process will be presented
in Obreja et al. (submitted).

Object HD-5103A, and its counterpart HD-5103B, deserve an special mention.
Since it has undergone a recent merger, its stellar content is mixed, and the
distributions are generally broader for any given radius, with many particles
shifting orbit and therefore lower particle densities attained in the
metallicity vs. radius plane.

\subsection{$\alpha$-elements}

A number of works in the literature, using different stellar samples, have
studied the abundance of $\alpha$-elements in the bulge, thin disc and thick
disc of the Milky Way \citep[e.g.][]{2003A&A...410..527B, 2005A&A...433..185B,
2006MNRAS.367.1329R, 2007A&A...474..763J, 2008A&A...484L..21M,
2010ApJ...721L..92R}. According to these works, the Galactic bulge and thick
disc are on average more enhanced in $\alpha$-elements than the thin disc,
suggesting a short formation time-scale for these two components. The
$\alpha$-elements are in fact mainly produced during Type II SNe on a short
time-scale, whereas iron is also produced by Type Ia SNe on a much longer
time-scale. An enhanced abundance in $\alpha$-elements can be then interpreted
as a signature of a short period of stellar formation, where SNe Ia have no
enough time to dilute the $\alpha$-elements produced by Type II supernovae.

Moreover, in the thin and thick disc the abundance of $\alpha$-elements show
tight and distinct trends with metallicity. In the thin disc, this trend has a
gentle slope whereas, for the thick disc, some authors
\citep{2003A&A...410..527B, 2005A&A...433..185B} have pointed out the presence
of a down-turn (or `knee'). According to these authors, for metallicity values
[Fe/H]$\lesssim-0.4$, stars in the thick disc have [$\alpha$/Fe] values higher
than in the thin disc. At [Fe/H]$\sim-0.4$, the [$\alpha$/Fe] trend has a
down-turn, so that it declines toward solar values merging with the thin disc
trend for [Fe/H]$\gtrsim-0.4$. In contrast, other authors
\citep[e.g.][]{2006MNRAS.367.1329R} doubt the existence of this `knee', but
agree in the different behaviour of thin and thick disc stars.

\begin{figure*}
\begin{center}
\includegraphics[width=0.8\textwidth]{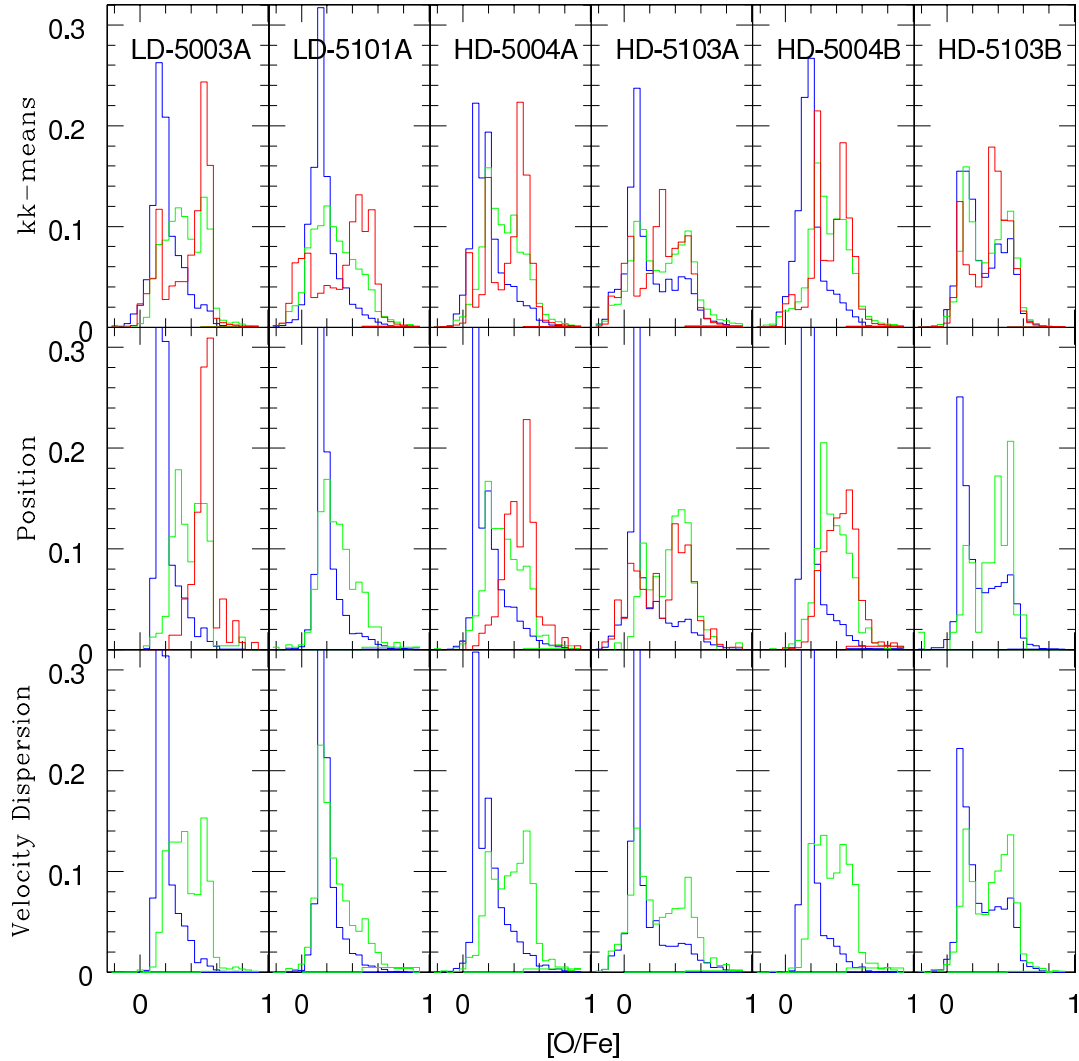}
\end{center}
\caption{Over-abundance distribution of stars in each galaxy component measured
as the oxygen ratio over iron.}
\label{mdf_alpha}
\end{figure*}

In Fig. \ref{mdf_alpha} we show the distribution of the abundance of
$\alpha$-elements represented by oxygen\footnote{Similar trends are found when
the abundance of $\alpha$-elements is represented by magnesium}. In general,
all samples of thick disc stars distribute around a median value ($\sim0.3$)
slightly higher than in the thin disc. For the location-based samples of stars
in the spheroid, we also find $\alpha$-enhanced abundances, with median [O/Fe]
values ($\sim0.45$) higher than in the thick disc. The spheroid samples
identified by `kk-means' have instead a bimodal distribution: there exists a
population of stars poor in $\alpha$-elements, as well as another population
enhanced in these elements. Such a bimodality could be due to the combined
action of two bulge formation mechanisms (classical bulge and pseudo-bulge),
and also to different origin of stars in the halo as recently pointed out by
\citet{2010A&A...511L..10N}. They suggest that low-$\alpha$ stars in the
Galactic halo probably were accreted from dwarf galaxies that had lower star
formation rates, while the high-$\alpha$ population is formed by the first
stars likely originated in a dissipative collapse of the proto-Galactic gas
cloud, or by ancient disc or bulge stars heated to the halo by merging
processes.

\begin{figure}
\begin{center}
\includegraphics[width=1.0\columnwidth]{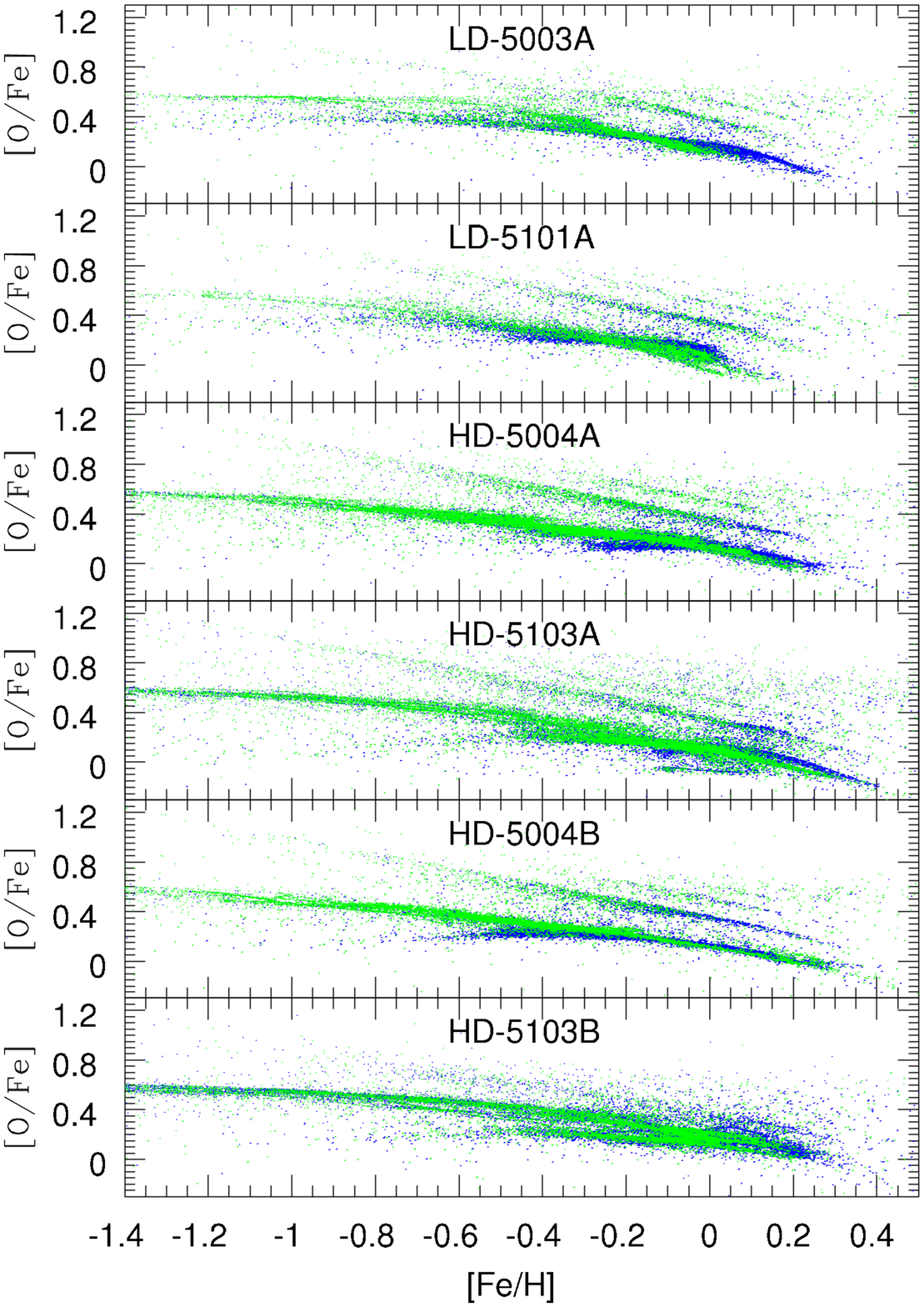}
\end{center}
\caption{
[O/Fe] versus [Fe/H] ratios for F and G stars in the thin and thick discs (blue
and green dots, respectively) of each simulated galaxy.}
\label{alpha}
\end{figure}

In Fig. \ref{alpha} we depict [O/Fe] versus [Fe/H] for disc stars in each
simulated galaxy. To compare our results with the observational trends reported
by \citet{2003A&A...410..527B, 2005A&A...433..185B}, the number of points in
Fig. \ref{alpha} are proportional to the fraction of F and G stars in each star
particle, computed from its age and metallicity assuming a Salpeter IMF. From
this figure we can see that at low metallicities the thick disc stars are
generally more $\alpha$-enhanced than those with the same metallicity in the
thin disc. This is particularly clear in objects LD-5003A and LD-5101A, located
in sparsely populated regions of the universe. In these objects, a `knee' is
found at roughly the same metallicity value ([Fe/H]$\sim-0.4$) as in the Milky
Way observations of \citet{2003A&A...410..527B, 2005A&A...433..185B}. We also
find for these objects that there are some old thin disc stars over-abundant in
oxygen. Although less clearly, such an $\alpha$-enhancement of the thick disc
low-metallicity stars seems to be also present in the objects HD-5004A,
HD-5103A and their counterparts, populating denser environments. We also note
that, except for HD-5103B, the plot of [O/Fe] versus [Fe/H] has secondary
branches containing stars of both disc components with high $\alpha$-abundance
and metallicity. These secondary curves are mainly populated by young thin disc
stars formed in an enriched medium and by old thick disc stars which formed
during a rapid episode. These old stars enriched the medium with
$\alpha$-elements and then they have been evolving at the same time that the
iron abundance has been increasing.

\section{Summary and conclusions}

We have presented a detailed analysis of the photometric and chemical
properties of seven simulations of disc galaxies in the $\Lambda$CDM scenario.
Each simulation is a cosmological zoom-in that includes high-resolution gas and
dark matter particles for the converging region that generates the main object.
Star formation is modelled through a Kennincutt-Schmidt-like law with a star
formation efficiency that implicitly accounts for energy feedback. Although the
addition of some explicit form of kinetic feedback could presumably lead to
simulated galaxy populations in even better agreement with observations
\citep{2009MNRAS.396.2332K}, in this paper we have preferred to test the
minimal conditions for the formation of realistic discs, thereby facilitating
the physical interpretation of our results. The methods for chemical feedback
and abundance-dependent cooling are also included. The simulations presented in
this paper correspond to four different objects selected from both sparsely
populated and crowded regions. In order to test the robustness of our results,
three of the seven runs consisted of variants with different SF efficiency or
resolution.

From the bulge-disc decomposition of their luminosity profiles, we find that
our simulated galaxies are disc dominated, with bulge sizes comparable to those
observed in late-type spirals. In addition they have also realistic kinematic
properties and lie in the region defined by the observational Tully-Fisher
relation.

In order to analyse the fine structure of these objects, we have performed a
dynamical decomposition of the galactic components based on the total binding
energy and the axial and radial components of the angular momentum, normalised
with the angular momentum expected for a particle in a circular orbit. We used
an unsupervised clustering algorithm which classifies stars by minimising the
sum of the distances of each particle to its cluster centre in the space
defined by these three parameters. In this way, we can identify stars from the
thin and thick disc, and also from the spheroid, which includes bulge and halo.
We will now discuss our main results for each component: spheroid, thin disc
and thick disc.

We find that the spheroidal component of the simulated galaxies is composed
mainly by old stars, with just a small contribution of stars younger than 5
Gyr. The bulk of stars formed in a rapid collapse at very early times.
Afterwards, the star formation turned off, and only residual star formation is
observed at low redshifts. The age and colour profiles for this component show
that the stellar content is distributed quite uniformly, as observed in real
galaxies, with mean ages almost constant along the galaxy. The only exception
is the simulated object HD-5103A and its less efficient SF counterpart HD-5103B
where, as a consequence of a recent major merger, an important star formation
episode took place at redshift $\sim0.3$. New stars were born in this event,
and populated mainly the spheroid, which shows an U-shaped colour profile and a
mean age profile younger than the other galaxies. The distribution of
metallicity and $\alpha$-elements for stars in the spheroid shows a clear
bimodality. Such a bimodality seems to be connected with the different origin
of stars in this component. The low-metallicity peak could be related to the
classical bulge, while the high-metallicity peak could be associated with a
pseudo-bulge formed from instabilities of the inner disc. A detailed analysis
of the formation mechanism and time evolution of the bulge component will be
presented in a forthcoming paper (Obreja et al. submitted).

The thin disc appears in our simulations as the youngest and most metal-rich
component, with roughly solar values for metallicity. The median age of stars
ranges from 3.8 to 6.7 Gyr, with a just small fraction older than 10 Gyr. This
is the only component with a non-negligible star formation rate ($\sim$ 1
M$_\odot$ yr$^{-1}$) at $z=0$. The radial distribution of ages and colours show
that new stars are forming in the thin disc according to the inside-out
paradigm. The U-shape of such profiles indicate
\citep[see][]{2009ApJ...705L.133M,2009MNRAS.398..591S} that stars have migrated
across the galaxy reaching high radial distances.

The thick disc in our simulations contains about 16 per cent of the total
stellar mass and has intermediate properties, in between those of the thin disc
and the spheroid. From its age and metallicity distributions we find that it is
populated by older and more metal-poor stars than in the thin disc, but younger
and more metal-rich stars than in the spheroid. The thick disc is not currently
forming stars, but its history of star formation is quieter than in the
spheroid, and extends in a long period of time, with a rate comparable to that
observed in the thin disc. The abundance distribution shows that the thick disc
is enhanced in $\alpha$-elements, but not as much as that found for the
spheroid.

Galaxies located in low-density regions of the universe have thick discs where,
for low-metallicity stars ([Fe/H]$\lesssim-0.4$), the abundance of
$\alpha$-elements is clearly enhanced as compared to thin disc stars of the
same metallicity. When fitting the vertical density profiles of RGB stars, we
find that objects in low-density regions have thick disc scale heights $z_h$
that do not change significantly along the disc. According to some authors
\citep{2009MNRAS.395..126I}, $z_h$ values roughly constant with the radial
distance favour the scenario in which the thickening results from internal
(non-merger-driven) instabilities in the thin disc. In the case of galaxies
located in high-density regions of the universe, we find instead a thick disc
scale height that changes with the galactocentric distance, as expected in a
merger-induced scenario for the formation of this component. For these
galaxies, the trend of [$\alpha$/Fe] versus [Fe/H] for the thin and thick discs
also present a different behaviour for [Fe/H]$\lesssim-0.4$. Nevertheless, now
the number of stars with very low metallicities in the thin disc is much
smaller than that found in the same component of galaxies in low-density
regions. Consequently, the trend of [$\alpha$/Fe] versus [Fe/H] for the thin
disc of high-density galaxies does not extend to very low metallicities. This
fact seems also point out a merger-driven scenario as the main mechanism for
the thick disc formation of galaxies in high-density regions. Indeed, the lack
of thin disc stars with very low metallicities (i.e., very old ages) for such
galaxies seems to favour a formation model in which an early thin disc was
almost completely disrupted by a violent event giving rise to the thick disc.
Afterwards, a new stellar thin disc was formed from the remaining and already
metal-enriched gas. In order to discern the role of the various mechanisms of
disc thickening a more detailed analysis of the galaxy evolution at different
redshifts evolution is needed. Such an analysis will be reported in a future
paper.

\section*{Acknowledgements}

FJMS would like to thank Brad Gibson and the Extragalactic Astrophysics group
at the Jeremiah Horrocks Institute for their kind hospitality and useful
discussions. We also thank the anonymous referee for his/her very useful
comments and suggestions. This work was partially supported by the DGES (Spain)
through the grants AYA2009-12792-C03-02 and AYA2009-12792-C03-03 from the
PNAyA, MICINN, Spain. We also thank the ASTROMADRID network (CAM
S2009/ESP-1496) from the Madrid regional government, and the
''Supercomputaci\'on y e-Ciencia'' Consolider-Ingenio 2010 project
(CSD2007-0050), MICINN, Spain. The Centro de Computaci\'on Cient\'ifica (UAM,
Spain) has provided computing facilities.

\bibliographystyle{mn2e}
\bibliography{bib_thickdisk}

\begin{thebibliography}{107}
\expandafter\ifx\csname natexlab\endcsname\relax\def\natexlab#1{#1}\fi

\bibitem[{{Abadi} {et~al}\mbox{.}(2003{\natexlab{a}}){Abadi}, {Navarro},
  {Steinmetz}, \& {Eke}}]{2003ApJ...591..499A}
{Abadi} M.~G., {Navarro} J.~F., {Steinmetz} M., {Eke} V.~R.,
  2003{\natexlab{a}}, \apj, 591, 499, arXiv:astro-ph/0211331

\bibitem[{{Abadi} {et~al}\mbox{.}(2003{\natexlab{b}}){Abadi}, {Navarro},
  {Steinmetz}, \& {Eke}}]{2003ApJ...597...21A}
{Abadi} M.~G., {Navarro} J.~F., {Steinmetz} M., {Eke} V.~R.,
  2003{\natexlab{b}}, \apj, 597, 21, arXiv:astro-ph/0212282

\bibitem[{{Agertz} {et~al}\mbox{.}(2011){Agertz}, {Teyssier}, \&
  {Moore}}]{2011MNRAS.410.1391A}
{Agertz} O., {Teyssier} R., {Moore} B., 2011, \mnras, 410, 1391, 1004.0005

\bibitem[{{Bakos} {et~al}\mbox{.}(2008){Bakos}, {Trujillo}, \&
  {Pohlen}}]{2008ApJ...683L.103B}
{Bakos} J., {Trujillo} I., {Pohlen} M., 2008, \apjl, 683, L103, 0807.2776

\bibitem[{Barber {et~al}\mbox{.}(1996)Barber, Dobkin, \&
  Huhdanpaa}]{barber96quickhull}
Barber C.~B., Dobkin D.~P., Huhdanpaa H., 1996, ACM Trans. Math. Softw., 22,
  469

\bibitem[{{Bensby} {et~al}\mbox{.}(2003){Bensby}, {Feltzing}, \&
  {Lundstr{\"o}m}}]{2003A&A...410..527B}
{Bensby} T., {Feltzing} S., {Lundstr{\"o}m} I., 2003, \aap, 410, 527

\bibitem[{{Bensby} {et~al}\mbox{.}(2005){Bensby}, {Feltzing}, {Lundstr{\"o}m},
  \& {Ilyin}}]{2005A&A...433..185B}
{Bensby} T., {Feltzing} S., {Lundstr{\"o}m} I., {Ilyin} I., 2005, \aap, 433,
  185, arXiv:astro-ph/0412132

\bibitem[{{Bigiel} {et~al}\mbox{.}(2010){Bigiel}, {Leroy}, {Walter}, {Blitz},
  {Brinks}, {de Blok}, \& {Madore}}]{2010AJ....140.1194B}
{Bigiel} F., {Leroy} A., {Walter} F., {Blitz} L., {Brinks} E., {de Blok}
  W.~J.~G., {Madore} B., 2010, \aj, 140, 1194, 1007.3498

\bibitem[{{Bournaud} {et~al}\mbox{.}(2009){Bournaud}, {Elmegreen}, \&
  {Martig}}]{2009ApJ...707L...1B}
{Bournaud} F., {Elmegreen} B.~G., {Martig} M., 2009, \apjl, 707, L1, 0910.3677

\bibitem[{{Brook} {et~al}\mbox{.}(2005){Brook}, {Gibson}, {Martel}, \&
  {Kawata}}]{2005ApJ...630..298B}
{Brook} C.~B., {Gibson} B.~K., {Martel} H., {Kawata} D., 2005, \apj, 630, 298,
  arXiv:astro-ph/0503273

\bibitem[{{Brook} {et~al}\mbox{.}(2004){Brook}, {Kawata}, {Gibson}, \&
  {Freeman}}]{2004ApJ...612..894B}
{Brook} C.~B., {Kawata} D., {Gibson} B.~K., {Freeman} K.~C., 2004, \apj, 612,
  894, arXiv:astro-ph/0405306

\bibitem[{{Brooks} {et~al}\mbox{.}(2009){Brooks}, {Governato}, {Quinn},
  {Brook}, \& {Wadsley}}]{2009ApJ...694..396B}
{Brooks} A.~M., {Governato} F., {Quinn} T., {Brook} C.~B., {Wadsley} J., 2009,
  \apj, 694, 396, 0812.0007

\bibitem[{{Bruzual} \& {Charlot}(2003)}]{2003MNRAS.344.1000B}
{Bruzual} G., {Charlot} S., 2003, \mnras, 344, 1000, arXiv:astro-ph/0309134

\bibitem[{{Burkert}(2009)}]{2009arXiv0908.1409B}
{Burkert} A., 2009, ArXiv e-prints, 0908.1409

\bibitem[{{Carollo} {et~al}\mbox{.}(2010){Carollo}, {Beers}, {Chiba}, {Norris},
  {Freeman}, {Lee}, {Ivezi{\'c}}, {Rockosi}, \& {Yanny}}]{2010ApJ...712..692C}
{Carollo} D. {et~al.}, 2010, \apj, 712, 692, 0909.3019

\bibitem[{{Catinella} {et~al}\mbox{.}(2006){Catinella}, {Giovanelli}, \&
  {Haynes}}]{2006ApJ...640..751C}
{Catinella} B., {Giovanelli} R., {Haynes} M.~P., 2006, \apj, 640, 751,
  arXiv:astro-ph/0512051

\bibitem[{{Collins} {et~al}\mbox{.}(2010){Collins}, {Chapman}, {Ibata},
  {Irwin}, {Rich}, {Ferguson}, {Lewis}, {Tanvir}, \&
  {Koch}}]{2010arXiv1010.5276C}
{Collins} M.~L.~M. {et~al.}, 2010, ArXiv e-prints, 1010.5276

\bibitem[{{Croft} {et~al}\mbox{.}(2009){Croft}, {Di Matteo}, {Springel}, \&
  {Hernquist}}]{2009MNRAS.400...43C}
{Croft} R.~A.~C., {Di Matteo} T., {Springel} V., {Hernquist} L., 2009, \mnras,
  400, 43, 0803.4003

\bibitem[{{Dalcanton} \& {Bernstein}(2002)}]{2002AJ....124.1328D}
{Dalcanton} J.~J., {Bernstein} R.~A., 2002, \aj, 124, 1328,
  arXiv:astro-ph/0207221

\bibitem[{{Dekel} \& {Birnboim}(2006)}]{2006MNRAS.368....2D}
{Dekel} A., {Birnboim} Y., 2006, \mnras, 368, 2, arXiv:astro-ph/0412300

\bibitem[{Dhillon {et~al}\mbox{.}(2004)Dhillon, Guan, \& Kulis}]{Dhillon04}
Dhillon I., Guan Y., Kulis B., 2004, A unified view of kernel k-means, spectral
  clustering and graph cuts. Tech. Rep. TR-04-25, University of Texas at
  Austin, Department of Computer Sciences

\bibitem[{{Dom{\'{\i}}nguez-Tenreiro}
  {et~al}\mbox{.}(1998){Dom{\'{\i}}nguez-Tenreiro}, {Tissera}, \&
  {S{\'a}iz}}]{1998ApJ...508L.123D}
{Dom{\'{\i}}nguez-Tenreiro} R., {Tissera} P.~B., {S{\'a}iz} A., 1998, \apjl,
  508, L123, arXiv:astro-ph/9809312

\bibitem[{{Edmunds} \& {Roy}(1993)}]{1993MNRAS.261L..17E}
{Edmunds} M.~G., {Roy} J.-R., 1993, \mnras, 261, L17

\bibitem[{{Fall}(1983)}]{1983IAUS..100..391F}
{Fall} S.~M., 1983, in IAU Symposium, Vol. 100, Internal Kinematics and
  Dynamics of Galaxies, {E.~Athanassoula}, ed., pp. 391--398

\bibitem[{{Fall} \& {Efstathiou}(1980)}]{1980MNRAS.193..189F}
{Fall} S.~M., {Efstathiou} G., 1980, \mnras, 193, 189

\bibitem[{{Fuhrmann}(1998)}]{1998A&A...338..161F}
{Fuhrmann} K., 1998, \aap, 338, 161

\bibitem[{{Gadotti}(2008)}]{2008MNRAS.384..420G}
{Gadotti} D.~A., 2008, \mnras, 384, 420, 0708.3870

\bibitem[{{Garnett} {et~al}\mbox{.}(1997){Garnett}, {Shields}, {Skillman},
  {Sagan}, \& {Dufour}}]{1997ApJ...489...63G}
{Garnett} D.~R., {Shields} G.~A., {Skillman} E.~D., {Sagan} S.~P., {Dufour}
  R.~J., 1997, \apj, 489, 63

\bibitem[{{Gavil{\'a}n} {et~al}\mbox{.}(2005){Gavil{\'a}n}, {Buell}, \&
  {Moll{\'a}}}]{2005A&A...432..861G}
{Gavil{\'a}n} M., {Buell} J.~F., {Moll{\'a}} M., 2005, \aap, 432, 861,
  arXiv:astro-ph/0411746

\bibitem[{{Gilmore} \& {Reid}(1983)}]{1983MNRAS.202.1025G}
{Gilmore} G., {Reid} N., 1983, \mnras, 202, 1025

\bibitem[{{Gilmore} {et~al}\mbox{.}(1995){Gilmore}, {Wyse}, \&
  {Jones}}]{1995AJ....109.1095G}
{Gilmore} G., {Wyse} R.~F.~G., {Jones} J.~B., 1995, \aj, 109, 1095,
  arXiv:astro-ph/9411116

\bibitem[{{Gonzalez-Perez} {et~al}\mbox{.}(2010){Gonzalez-Perez}, {Castander},
  \& {Kauffmann}}]{2010MNRAS.tmp.1828G}
{Gonzalez-Perez} V., {Castander} F.~J., {Kauffmann} G., 2010, \mnras, 1828,
  1008.2354

\bibitem[{{Governato} {et~al}\mbox{.}(2009){Governato}, {Brook}, {Brooks},
  {Mayer}, {Willman}, {Jonsson}, {Stilp}, {Pope}, {Christensen}, {Wadsley}, \&
  {Quinn}}]{2009MNRAS.398..312G}
{Governato} F. {et~al.}, 2009, \mnras, 398, 312, 0812.0379

\bibitem[{{Governato} {et~al}\mbox{.}(2007){Governato}, {Willman}, {Mayer},
  {Brooks}, {Stinson}, {Valenzuela}, {Wadsley}, \&
  {Quinn}}]{2007MNRAS.374.1479G}
{Governato} F., {Willman} B., {Mayer} L., {Brooks} A., {Stinson} G.,
  {Valenzuela} O., {Wadsley} J., {Quinn} T., 2007, \mnras, 374, 1479,
  arXiv:astro-ph/0602351

\bibitem[{{Graham} \& {Worley}(2008)}]{2008MNRAS.388.1708G}
{Graham} A.~W., {Worley} C.~C., 2008, \mnras, 388, 1708, 0805.3565

\bibitem[{{Grevesse} \& {Sauval}(1998)}]{1998SSRv...85..161G}
{Grevesse} N., {Sauval} A.~J., 1998, \ssr, 85, 161

\bibitem[{{Guedes} {et~al}\mbox{.}(2011){Guedes}, {Callegari}, {Madau}, \&
  {Mayer}}]{2011arXiv1103.6030G}
{Guedes} J., {Callegari} S., {Madau} P., {Mayer} L., 2011, ArXiv e-prints,
  1103.6030

\bibitem[{{Ibata} {et~al}\mbox{.}(2009){Ibata}, {Mouhcine}, \&
  {Rejkuba}}]{2009MNRAS.395..126I}
{Ibata} R., {Mouhcine} M., {Rejkuba} M., 2009, \mnras, 395, 126, 0903.4209

\bibitem[{{Ivezi{\'c}} {et~al}\mbox{.}(2008){Ivezi{\'c}}, {Sesar}, {Juri{\'c}},
  {Bond}, {Dalcanton}, {Rockosi}, {Yanny}, {Newberg}, {Beers}, {Allende
  Prieto}, {Wilhelm}, {Lee}, {Sivarani}, {Norris}, {Bailer-Jones}, {Re
  Fiorentin}, {Schlegel}, {Uomoto}, {Lupton}, {Knapp}, {Gunn}, {Covey},
  {Smith}, {Miknaitis}, {Doi}, {Tanaka}, {Fukugita}, {Kent}, {Finkbeiner},
  {Munn}, {Pier}, {Quinn}, {Hawley}, {Anderson}, {Kiuchi}, {Chen}, {Bushong},
  {Sohi}, {Haggard}, {Kimball}, {Barentine}, {Brewington}, {Harvanek},
  {Kleinman}, {Krzesinski}, {Long}, {Nitta}, {Snedden}, {Lee}, {Harris},
  {Brinkmann}, {Schneider}, \& {York}}]{2008ApJ...684..287I}
{Ivezi{\'c}} {\v Z}. {et~al.}, 2008, \apj, 684, 287, 0804.3850

\bibitem[{{Iwamoto} {et~al}\mbox{.}(1999){Iwamoto}, {Brachwitz}, {Nomoto},
  {Kishimoto}, {Umeda}, {Hix}, \& {Thielemann}}]{1999ApJS..125..439I}
{Iwamoto} K., {Brachwitz} F., {Nomoto} K., {Kishimoto} N., {Umeda} H., {Hix}
  W.~R., {Thielemann} F.-K., 1999, \apjs, 125, 439, arXiv:astro-ph/0002337

\bibitem[{{Jablonka} {et~al}\mbox{.}(2007){Jablonka}, {Gorgas}, \&
  {Goudfrooij}}]{2007A&A...474..763J}
{Jablonka} P., {Gorgas} J., {Goudfrooij} P., 2007, \aap, 474, 763, 0707.0561

\bibitem[{{Juri{\'c}} {et~al}\mbox{.}(2008){Juri{\'c}}, {Ivezi{\'c}}, {Brooks},
  {Lupton}, {Schlegel}, {Finkbeiner}, {Padmanabhan}, {Bond}, {Sesar},
  {Rockosi}, {Knapp}, {Gunn}, {Sumi}, {Schneider}, {Barentine}, {Brewington},
  {Brinkmann}, {Fukugita}, {Harvanek}, {Kleinman}, {Krzesinski}, {Long},
  {Neilsen}, {Nitta}, {Snedden}, \& {York}}]{2008ApJ...673..864J}
{Juri{\'c}} M. {et~al.}, 2008, \apj, 673, 864, arXiv:astro-ph/0510520

\bibitem[{Karatzoglou {et~al}\mbox{.}(2004)Karatzoglou, Smola, \&
  Hornik}]{Karatzoglou04kernlab}
Karatzoglou A., Smola A., Hornik K., 2004, Journal of Statistical Software, 1

\bibitem[{{Kere{\v s}} {et~al}\mbox{.}(2009){Kere{\v s}}, {Katz}, {Dav{\'e}},
  {Fardal}, \& {Weinberg}}]{2009MNRAS.396.2332K}
{Kere{\v s}} D., {Katz} N., {Dav{\'e}} R., {Fardal} M., {Weinberg} D.~H., 2009,
  \mnras, 396, 2332, 0901.1880

\bibitem[{{Kere{\v s}} {et~al}\mbox{.}(2005){Kere{\v s}}, {Katz}, {Weinberg},
  \& {Dav{\'e}}}]{2005MNRAS.363....2K}
{Kere{\v s}} D., {Katz} N., {Weinberg} D.~H., {Dav{\'e}} R., 2005, \mnras, 363,
  2, arXiv:astro-ph/0407095

\bibitem[{{Kormendy} \& {Kennicutt}(2004)}]{2004ARA&A..42..603K}
{Kormendy} J., {Kennicutt}, Jr. R.~C., 2004, \araa, 42, 603,
  arXiv:astro-ph/0407343

\bibitem[{{Kravtsov}(2003)}]{2003ApJ...590L...1K}
{Kravtsov} A.~V., 2003, \apjl, 590, L1, arXiv:astro-ph/0303240

\bibitem[{{Laurikainen} {et~al}\mbox{.}(2007){Laurikainen}, {Salo}, {Buta}, \&
  {Knapen}}]{2007MNRAS.381..401L}
{Laurikainen} E., {Salo} H., {Buta} R., {Knapen} J.~H., 2007, \mnras, 381, 401,
  arXiv:astro-ph/0702434

\bibitem[{{Leroy} {et~al}\mbox{.}(2008){Leroy}, {Walter}, {Brinks}, {Bigiel},
  {de Blok}, {Madore}, \& {Thornley}}]{2008AJ....136.2782L}
{Leroy} A.~K., {Walter} F., {Brinks} E., {Bigiel} F., {de Blok} W.~J.~G.,
  {Madore} B., {Thornley} M.~D., 2008, \aj, 136, 2782, 0810.2556

\bibitem[{{Luck} {et~al}\mbox{.}(2006){Luck}, {Kovtyukh}, \&
  {Andrievsky}}]{2006AJ....132..902L}
{Luck} R.~E., {Kovtyukh} V.~V., {Andrievsky} S.~M., 2006, \aj, 132, 902

\bibitem[{{MacArthur} {et~al}\mbox{.}(2009){MacArthur}, {Gonz{\'a}lez}, \&
  {Courteau}}]{2009MNRAS.395...28M}
{MacArthur} L.~A., {Gonz{\'a}lez} J.~J., {Courteau} S., 2009, \mnras, 395, 28,
  0901.4135

\bibitem[{{Mart{\'{\i}}nez-Serrano}
  {et~al}\mbox{.}(2009){Mart{\'{\i}}nez-Serrano}, {Serna},
  {Dom{\'e}nech-Moral}, \& {Dom{\'{\i}}nguez-Tenreiro}}]{2009ApJ...705L.133M}
{Mart{\'{\i}}nez-Serrano} F.~J., {Serna} A., {Dom{\'e}nech-Moral} M.,
  {Dom{\'{\i}}nguez-Tenreiro} R., 2009, \apjl, 705, L133, 0906.1118

\bibitem[{{Mart{\'{\i}}nez-Serrano}
  {et~al}\mbox{.}(2008){Mart{\'{\i}}nez-Serrano}, {Serna},
  {Dom{\'{\i}}nguez-Tenreiro}, \& {Moll{\'a}}}]{2008MNRAS.388...39M}
{Mart{\'{\i}}nez-Serrano} F.~J., {Serna} A., {Dom{\'{\i}}nguez-Tenreiro} R.,
  {Moll{\'a}} M., 2008, \mnras, 388, 39, 0804.3766

\bibitem[{{Mel{\'e}ndez} {et~al}\mbox{.}(2008){Mel{\'e}ndez}, {Asplund},
  {Alves-Brito}, {Cunha}, {Barbuy}, {Bessell}, {Chiappini}, {Freeman},
  {Ram{\'{\i}}rez}, {Smith}, \& {Yong}}]{2008A&A...484L..21M}
{Mel{\'e}ndez} J. {et~al.}, 2008, \aap, 484, L21, 0804.4124

\bibitem[{{Navarro}(1998)}]{1998astro.ph..7084N}
{Navarro} J.~F., 1998, ArXiv Astrophysics e-prints, arXiv:astro-ph/9807084

\bibitem[{{Navarro} {et~al}\mbox{.}(2011){Navarro}, {Abadi}, {Venn}, {Freeman},
  \& {Anguiano}}]{2011MNRAS.412.1203N}
{Navarro} J.~F., {Abadi} M.~G., {Venn} K.~A., {Freeman} K.~C., {Anguiano} B.,
  2011, \mnras, 412, 1203, 1009.0020

\bibitem[{{Navarro} \& {White}(1994)}]{1994MNRAS.267..401N}
{Navarro} J.~F., {White} S.~D.~M., 1994, \mnras, 267, 401

\bibitem[{{Nissen} \& {Schuster}(2010)}]{2010A&A...511L..10N}
{Nissen} P.~E., {Schuster} W.~J., 2010, \aap, 511, L10+, 1002.4514

\bibitem[{{Obreja} {et~al}\mbox{.}(2009){Obreja}, {Dom\'inguez-Tenreiro}, {Mart\'inez-Serrano},
  {Dom\'enech-Moral}, \& {Serna}}]{Obrejasub}
{Obreja} A., {Dom\'inguez-Tenreiro} R., {Mart\'inez-Serrano} F., {Dom\'enech-Moral} M., {Serna} A., 2012,
  \mnras, submitted

\bibitem[{{Oohama} {et~al}\mbox{.}(2009){Oohama}, {Okamura}, {Fukugita},
  {Yasuda}, \& {Nakamura}}]{2009ApJ...705..245O}
{Oohama} N., {Okamura} S., {Fukugita} M., {Yasuda} N., {Nakamura} O., 2009,
  \apj, 705, 245, 0908.4312

\bibitem[{{Pancino} {et~al}\mbox{.}(2010){Pancino}, {Carrera}, {Rossetti}, \&
  {Gallart}}]{2010A&A...511A..56P}
{Pancino} E., {Carrera} R., {Rossetti} E., {Gallart} C., 2010, \aap, 511, A56+,
  0910.0723

\bibitem[{{Peng} {et~al}\mbox{.}(2010){Peng}, {Ho}, {Impey}, \&
  {Rix}}]{2010AJ....139.2097P}
{Peng} C.~Y., {Ho} L.~C., {Impey} C.~D., {Rix} H.-W., 2010, \aj, 139, 2097,
  0912.0731

\bibitem[{{Power} \& {Knebe}(2006)}]{2006MNRAS.370..691P}
{Power} C., {Knebe} A., 2006, \mnras, 370, 691, arXiv:astro-ph/0512281

\bibitem[{{Prunet} {et~al}\mbox{.}(2008){Prunet}, {Pichon}, {Aubert},
  {Pogosyan}, {Teyssier}, \& {Gottloeber}}]{2008ApJS..178..179P}
{Prunet} S., {Pichon} C., {Aubert} D., {Pogosyan} D., {Teyssier} R.,
  {Gottloeber} S., 2008, \apjs, 178, 179, 0804.3536

\bibitem[{{Quinn} {et~al}\mbox{.}(1993){Quinn}, {Hernquist}, \&
  {Fullagar}}]{1993ApJ...403...74Q}
{Quinn} P.~J., {Hernquist} L., {Fullagar} D.~P., 1993, \apj, 403, 74

\bibitem[{{Reddy} {et~al}\mbox{.}(2006){Reddy}, {Lambert}, \& {Allende
  Prieto}}]{2006MNRAS.367.1329R}
{Reddy} B.~E., {Lambert} D.~L., {Allende Prieto} C., 2006, \mnras, 367, 1329,
  arXiv:astro-ph/0512505

\bibitem[{{Robin} {et~al}\mbox{.}(2003){Robin}, {Reyl{\'e}}, {Derri{\`e}re}, \&
  {Picaud}}]{2003A&A...409..523R}
{Robin} A.~C., {Reyl{\'e}} C., {Derri{\`e}re} S., {Picaud} S., 2003, \aap, 409,
  523

\bibitem[{{Ro{\v s}kar} {et~al}\mbox{.}(2008){Ro{\v s}kar}, {Debattista},
  {Quinn}, {Stinson}, \& {Wadsley}}]{2008ApJ...684L..79R}
{Ro{\v s}kar} R., {Debattista} V.~P., {Quinn} T.~R., {Stinson} G.~S., {Wadsley}
  J., 2008, \apjl, 684, L79, 0808.0206

\bibitem[{{Ruchti} {et~al}\mbox{.}(2010){Ruchti}, {Fulbright}, {Wyse},
  {Gilmore}, {Bienaym{\'e}}, {Binney}, {Bland-Hawthorn}, {Campbell}, {Freeman},
  {Gibson}, {Grebel}, {Helmi}, {Munari}, {Navarro}, {Parker}, {Reid},
  {Seabroke}, {Siebert}, {Siviero}, {Steinmetz}, {Watson}, {Williams}, \&
  {Zwitter}}]{2010ApJ...721L..92R}
{Ruchti} G.~R. {et~al.}, 2010, \apjl, 721, L92, 1008.3828

\bibitem[{{Ruiz-Lapuente} {et~al}\mbox{.}(2000){Ruiz-Lapuente}, {Blinnikov},
  {Canal}, {Mendez}, {Sorokina}, {Visco}, \& {Walton}}]{2000MmSAI..71..435R}
{Ruiz-Lapuente} P., {Blinnikov} S., {Canal} R., {Mendez} J., {Sorokina} E.,
  {Visco} A., {Walton} N., 2000, \memsai, 71, 435

\bibitem[{{S{\'a}iz} {et~al}\mbox{.}(2001){S{\'a}iz},
  {Dom{\'{\i}}nguez-Tenreiro}, {Tissera}, \& {Courteau}}]{2001MNRAS.325..119S}
{S{\'a}iz} A., {Dom{\'{\i}}nguez-Tenreiro} R., {Tissera} P.~B., {Courteau} S.,
  2001, \mnras, 325, 119, arXiv:astro-ph/0102011

\bibitem[{{Sales} {et~al}\mbox{.}(2009){Sales}, {Helmi}, {Abadi}, {Brook},
  {G{\'o}mez}, {Ro{\v s}kar}, {Debattista}, {House}, {Steinmetz}, \&
  {Villalobos}}]{2009MNRAS.400L..61S}
{Sales} L.~V. {et~al.}, 2009, \mnras, 400, L61, 0909.3858

\bibitem[{{Salpeter}(1955)}]{1955ApJ...121..161S}
{Salpeter} E.~E., 1955, \apj, 121, 161

\bibitem[{{S{\'a}nchez-Bl{\'a}zquez}
  {et~al}\mbox{.}(2009){S{\'a}nchez-Bl{\'a}zquez}, {Courty}, {Gibson}, \&
  {Brook}}]{2009MNRAS.398..591S}
{S{\'a}nchez-Bl{\'a}zquez} P., {Courty} S., {Gibson} B.~K., {Brook} C.~B.,
  2009, \mnras, 398, 591, 0905.4579

\bibitem[{{Scannapieco} {et~al}\mbox{.}(2010){Scannapieco}, {Gadotti},
  {Jonsson}, \& {White}}]{2010MNRAS.407L..41S}
{Scannapieco} C., {Gadotti} D.~A., {Jonsson} P., {White} S.~D.~M., 2010,
  \mnras, 407, L41, 1001.4890

\bibitem[{{Scannapieco} \& {Tissera}(2003)}]{2003MNRAS.338..880S}
{Scannapieco} C., {Tissera} P.~B., 2003, \mnras, 338, 880,
  arXiv:astro-ph/0208538

\bibitem[{{Scannapieco} {et~al}\mbox{.}(2008){Scannapieco}, {Tissera}, {White},
  \& {Springel}}]{2008MNRAS.389.1137S}
{Scannapieco} C., {Tissera} P.~B., {White} S.~D.~M., {Springel} V., 2008,
  \mnras, 389, 1137, 0804.3795

\bibitem[{{Scannapieco} {et~al}\mbox{.}(2011{\natexlab{a}}){Scannapieco},
  {Wadepuhl}, {Parry}, {Navarro}, {Jenkins}, {Springel}, {Teyssier}, {Carlson},
  {Couchman}, {Crain}, {Dalla Vecchia}, {Frenk}, {Kobayashi}, {Monaco},
  {Murante}, {Okamoto}, {Quinn}, {Schaye}, {Stinson}, {Theuns}, {Wadsley},
  {White}, \& {Woods}}]{2011arXiv1112.0315S}
{Scannapieco} C. {et~al.}, 2011{\natexlab{a}}, ArXiv e-prints, 1112.0315

\bibitem[{{Scannapieco} {et~al}\mbox{.}(2009){Scannapieco}, {White},
  {Springel}, \& {Tissera}}]{2009MNRAS.396..696S}
{Scannapieco} C., {White} S.~D.~M., {Springel} V., {Tissera} P.~B., 2009,
  \mnras, 396, 696, 0812.0976

\bibitem[{{Scannapieco} {et~al}\mbox{.}(2011{\natexlab{b}}){Scannapieco},
  {White}, {Springel}, \& {Tissera}}]{2011MNRAS.417..154S}
{Scannapieco} C., {White} S.~D.~M., {Springel} V., {Tissera} P.~B.,
  2011{\natexlab{b}}, \mnras, 417, 154, 1105.0680

\bibitem[{Sch\"{o}lkopf {et~al}\mbox{.}(1998)Sch\"{o}lkopf, Smola, \&
  M\"{u}ller}]{Scholkopf1998}
Sch\"{o}lkopf B., Smola A., M\"{u}ller K.-R., 1998, Neural Comput., 10, 1299

\bibitem[{{Sch{\"o}nrich} \&
  {Binney}(2009{\natexlab{a}})}]{2009MNRAS.396..203S}
{Sch{\"o}nrich} R., {Binney} J., 2009{\natexlab{a}}, \mnras, 396, 203,
  0809.3006

\bibitem[{{Sch{\"o}nrich} \&
  {Binney}(2009{\natexlab{b}})}]{2009MNRAS.399.1145S}
{Sch{\"o}nrich} R., {Binney} J., 2009{\natexlab{b}}, \mnras, 399, 1145,
  0907.1899

\bibitem[{{Schuster} {et~al}\mbox{.}(2006){Schuster}, {Moitinho},
  {M{\'a}rquez}, {Parrao}, \& {Covarrubias}}]{2006A&A...445..939S}
{Schuster} W.~J., {Moitinho} A., {M{\'a}rquez} A., {Parrao} L., {Covarrubias}
  E., 2006, \aap, 445, 939, arXiv:astro-ph/0510313

\bibitem[{{Serna} {et~al}\mbox{.}(2003){Serna}, {Dom{\'{\i}}nguez-Tenreiro}, \&
  {S{\'a}iz}}]{2003ApJ...597..878S}
{Serna} A., {Dom{\'{\i}}nguez-Tenreiro} R., {S{\'a}iz} A., 2003, \apj, 597,
  878, arXiv:astro-ph/0307312

\bibitem[{{Seth} {et~al}\mbox{.}(2005){Seth}, {Dalcanton}, \& {de
  Jong}}]{2005AJ....130.1574S}
{Seth} A.~C., {Dalcanton} J.~J., {de Jong} R.~S., 2005, \aj, 130, 1574,
  arXiv:astro-ph/0506117

\bibitem[{{Shaver} {et~al}\mbox{.}(1983){Shaver}, {McGee}, {Newton}, {Danks},
  \& {Pottasch}}]{1983MNRAS.204...53S}
{Shaver} P.~A., {McGee} R.~X., {Newton} L.~M., {Danks} A.~C., {Pottasch} S.~R.,
  1983, \mnras, 204, 53

\bibitem[{{Shen} {et~al}\mbox{.}(2003){Shen}, {Mo}, {White}, {Blanton},
  {Kauffmann}, {Voges}, {Brinkmann}, \& {Csabai}}]{2003MNRAS.343..978S}
{Shen} S., {Mo} H.~J., {White} S.~D.~M., {Blanton} M.~R., {Kauffmann} G.,
  {Voges} W., {Brinkmann} J., {Csabai} I., 2003, \mnras, 343, 978,
  arXiv:astro-ph/0301527

\bibitem[{{Siegel} {et~al}\mbox{.}(2002){Siegel}, {Majewski}, {Reid}, \&
  {Thompson}}]{2002ApJ...578..151S}
{Siegel} M.~H., {Majewski} S.~R., {Reid} I.~N., {Thompson} I.~B., 2002, \apj,
  578, 151, arXiv:astro-ph/0206323

\bibitem[{{Sommer-Larsen} {et~al}\mbox{.}(2003){Sommer-Larsen}, {G{\"o}tz}, \&
  {Portinari}}]{2003ApJ...596...47S}
{Sommer-Larsen} J., {G{\"o}tz} M., {Portinari} L., 2003, \apj, 596, 47,
  arXiv:astro-ph/0204366

\bibitem[{{Soubiran} {et~al}\mbox{.}(2003){Soubiran}, {Bienaym{\'e}}, \&
  {Siebert}}]{2003A&A...398..141S}
{Soubiran} C., {Bienaym{\'e}} O., {Siebert} A., 2003, \aap, 398, 141,
  arXiv:astro-ph/0210628

\bibitem[{{Springob} {et~al}\mbox{.}(2007){Springob}, {Masters}, {Haynes},
  {Giovanelli}, \& {Marinoni}}]{2007ApJS..172..599S}
{Springob} C.~M., {Masters} K.~L., {Haynes} M.~P., {Giovanelli} R., {Marinoni}
  C., 2007, \apjs, 172, 599

\bibitem[{{Stinson} {et~al}\mbox{.}(2010){Stinson}, {Bailin}, {Couchman},
  {Wadsley}, {Shen}, {Nickerson}, {Brook}, \& {Quinn}}]{2010MNRAS.408..812S}
{Stinson} G.~S., {Bailin} J., {Couchman} H., {Wadsley} J., {Shen} S.,
  {Nickerson} S., {Brook} C., {Quinn} T., 2010, \mnras, 408, 812

\bibitem[{{Talbot} \& {Arnett}(1973)}]{1973ApJ...186...51T}
{Talbot}, Jr. R.~J., {Arnett} W.~D., 1973, \apj, 186, 51

\bibitem[{{Torrey} {et~al}\mbox{.}(2011){Torrey}, {Vogelsberger}, {Sijacki},
  {Springel}, \& {Hernquist}}]{2011arXiv1110.5635T}
{Torrey} P., {Vogelsberger} M., {Sijacki} D., {Springel} V., {Hernquist} L.,
  2011, ArXiv e-prints, 1110.5635

\bibitem[{{van Zee} {et~al}\mbox{.}(1998){van Zee}, {Salzer}, {Haynes},
  {O'Donoghue}, \& {Balonek}}]{1998AJ....116.2805V}
{van Zee} L., {Salzer} J.~J., {Haynes} M.~P., {O'Donoghue} A.~A., {Balonek}
  T.~J., 1998, \aj, 116, 2805, arXiv:astro-ph/9808315

\bibitem[{{Villalobos} \& {Helmi}(2008)}]{2008MNRAS.391.1806V}
{Villalobos} {\'A}., {Helmi} A., 2008, \mnras, 391, 1806

\bibitem[{{Wada} \& {Norman}(2007)}]{2007ApJ...660..276W}
{Wada} K., {Norman} C.~A., 2007, \apj, 660, 276, arXiv:astro-ph/0701595

\bibitem[{{Walker} {et~al}\mbox{.}(1996){Walker}, {Mihos}, \&
  {Hernquist}}]{1996ApJ...460..121W}
{Walker} I.~R., {Mihos} J.~C., {Hernquist} L., 1996, \apj, 460, 121,
  arXiv:astro-ph/9510052

\bibitem[{{Weinzirl} {et~al}\mbox{.}(2009){Weinzirl}, {Jogee}, {Khochfar},
  {Burkert}, \& {Kormendy}}]{2009ApJ...696..411W}
{Weinzirl} T., {Jogee} S., {Khochfar} S., {Burkert} A., {Kormendy} J., 2009,
  \apj, 696, 411, 0807.0040

\bibitem[{{White}(1984)}]{1984ApJ...286...38W}
{White} S.~D.~M., 1984, \apj, 286, 38

\bibitem[{{Woosley} \& {Weaver}(1995)}]{1995ApJS..101..181W}
{Woosley} S.~E., {Weaver} T.~A., 1995, \apjs, 101, 181

\bibitem[{{Yoachim} \& {Dalcanton}(2005)}]{2005ApJ...624..701Y}
{Yoachim} P., {Dalcanton} J.~J., 2005, \apj, 624, 701, arXiv:astro-ph/0501394

\bibitem[{{Yoachim} \& {Dalcanton}(2006)}]{2006AJ....131..226Y}
{Yoachim} P., {Dalcanton} J.~J., 2006, \aj, 131, 226, arXiv:astro-ph/0508460

\bibitem[{{Yoachim} \& {Dalcanton}(2008)}]{2008ApJ...683..707Y}
{Yoachim} P., {Dalcanton} J.~J., 2008, \apj, 683, 707, 0805.4197

\bibitem[{{Yong} {et~al}\mbox{.}(2005){Yong}, {Carney}, \& {Teixera de
  Almeida}}]{2005AJ....130..597Y}
{Yong} D., {Carney} B.~W., {Teixera de Almeida} M.~L., 2005, \aj, 130, 597,
  arXiv:astro-ph/0504193

\bibitem[{{Zaritsky} {et~al}\mbox{.}(1994){Zaritsky}, {Kennicutt}, \&
  {Huchra}}]{1994ApJ...420...87Z}
{Zaritsky} D., {Kennicutt}, Jr. R.~C., {Huchra} J.~P., 1994, \apj, 420, 87

\bibitem[{{Zoccali} {et~al}\mbox{.}(2008){Zoccali}, {Hill}, {Lecureur},
  {Barbuy}, {Renzini}, {Minniti}, {G{\'o}mez}, \&
  {Ortolani}}]{2008A&A...486..177Z}
{Zoccali} M., {Hill} V., {Lecureur} A., {Barbuy} B., {Renzini} A., {Minniti}
  D., {G{\'o}mez} A., {Ortolani} S., 2008, \aap, 486, 177, 0805.1218

\end{thebibliography}

\label{lastpage}

\end{document}